\documentclass{article}
\usepackage{slashed,braket,bm,latexsym,amsmath,amssymb,amsfonts,fancyhdr,color,graphicx,multirow,slashed,cite}
\usepackage{array}
\usepackage{booktabs}
\usepackage[a4paper,bottom=3cm,top=2.5cm,head=0mm,width=17cm,dvipdfm]{geometry}
\usepackage[usenames,dvipsnames,svgnames,table]{xcolor}
\usepackage[colorlinks=true,
            linkcolor=blue,
            urlcolor=blue,
            citecolor=green,          
						bookmarks=true,
						bookmarksnumbered=true,
						breaklinks=true,
						pdfpagemode=FullScreen,
						pdfstartview=FitBH]{hyperref}
\usepackage[dotinlabels]{titletoc}
\usepackage{titlesec}
\usepackage{authblk,ulem}

\numberwithin{equation}{section}
\allowdisplaybreaks[4]

\titlelabel{\thetitle.\quad \hspace{-0.8em}}
\titlecontents{section}
              [1.5em]
              {\vspace{4mm} \large \bf}
              {\contentslabel{1em}}
              {\hspace*{-1em}}
              {\titlerule*[.5pc]{.}\contentspage}
\titlecontents{subsection}
              [3.5em]
              {\vspace{2mm}}
              {\contentslabel{1.8em}}
              {\hspace*{.3em}}
              {\titlerule*[.5pc]{.}\contentspage}
\titlecontents{subsubsection}
              [5.5em]
              {\vspace{2mm}}
              {\contentslabel{2.5em}}
              {\hspace*{.3em}}
              {\titlerule*[.5pc]{.}\contentspage}

\newcommand{\titledef}{Relativistic Atomic Effects of Dark Matter Electron Scattering} % Insert Title here!!!
\hypersetup{ pdfauthor = {Shao-Feng Ge},
	     pdftitle = {\titledef}, % Insert title here!!!
	     pdfsubject = {}, % Insert subject here!!!
             pdfkeywords = {}, % Insert keywords here!!!
	     pdfcreator = {LaTeX with hyperref package},
	     pdfproducer = {dvips + ps2pdf} }

\newcommand{\tr}{\mbox{Tr}}

\definecolor{gesfblack}{rgb}{0,0,0}

\definecolor{gesfblue}{rgb}{0.08,0.42,0.76}

\definecolor{gesfgreen}{rgb}{0,1,0}

\definecolor{gesfgrey}{rgb}{0.5,0.5,0.5}

\definecolor{gesflanse}{rgb}{0.00,0.50,0.50}

\definecolor{gesfpurple}{rgb}{0.47,0.19,0.42}

\definecolor{gesfred}{rgb}{1,0,0}

\definecolor{gesfwhite}{rgb}{1,1,1}

\definecolor{gesfyellow}{rgb}{0.7,0.4,0.3}

\newcommand{\gsec}[1]{{\hypersetup{linkcolor=red}Sec.\,\ref{#1}\hypersetup{linkcolor=blue}}}
\newcommand{\gapp}[1]{{\hypersetup{linkcolor=red}App.\,\ref{#1}\hypersetup{linkcolor=blue}}}
\newcommand{\geqn}[1]{\hypersetup{linkcolor=blue}Eq.\,(\ref{#1})\hypersetup{linkcolor=blue}}
\newcommand{\gfig}[1]{{\hypersetup{linkcolor=violet}Fig.\,\ref{#1}\hypersetup{linkcolor=blue}}}
\newcommand{\gtab}[1]{{\hypersetup{linkcolor=gesflanse}Table~\ref{#1}\hypersetup{linkcolor=blue}}}

\newcommand{\dFs}[1]{\frac{d^3 #1}{(2\pi)^3 2 E_{ #1}}}
\newcommand{\dFp}[1]{\frac{d^3 #1}{(2\pi)^3 \sqrt{2 E_{ #1}}}}

\definecolor{Orange}{cmyk}{0,0.61,0.87,0}
\definecolor{JungleGreen}{cmyk}{0.99,0,0.52,0}
\definecolor{OliveGreen}{cmyk}{0.64,0,0.95,0.40}
\definecolor{Brown}{cmyk}{0,0.81,1,0.60}
\definecolor{RoyalBlue}{cmyk}{0.71,0.53,0,0.12}
\definecolor{Gray}{cmyk}{0,0,0,0.40}
\definecolor{LightPink}{cmyk}{0.0,0.25,0,0}
\definecolor{LLightPink}{cmyk}{0.0,0.10,0,0}
\definecolor{LightBlue}{cmyk}{0.25,0,0,0}
\definecolor{LightGray}{cmyk}{0,0,0,0.2}

\usepackage{subfigure}
\newcommand{\bee}{\begin{equation}}
\newcommand{\ene}{\end{equation}}
\newcommand{\bea}{\begin{eqnarray}}
\newcommand{\ena}{\end{eqnarray}}

\begin{document}
\fontsize{12pt}{14pt}\selectfont

\title{%\begin{flushright}
       %\mbox{\normalsize IPMU18-xxxx}
       %\end{flushright}
			 %\vskip 20pt
       \Large \bf \titledef} % Insert title here!!!
\author[1,2]{{\large Shao-Feng Ge} \footnote{\href{mailto:gesf@sjtu.edu.cn}{gesf@sjtu.edu.cn} (Corresponding Author)}}
\affil[1]{State Key Laboratory of Dark Matter Physics, Tsung-Dao Lee Institute \& School of Physics and Astronomy, Shanghai Jiao Tong University, Shanghai 200240, China}
\affil[2]{Key Laboratory for Particle Astrophysics and Cosmology (MOE) \& Shanghai Key Laboratory for Particle Physics and Cosmology, Shanghai Jiao Tong University, Shanghai 200240, China}
\author[1,2,3]{{\large Jie Sheng}
\affil[3]{
Kavli IPMU (WPI), UTIAS, University of Tokyo, Kashiwa, 277-8583, Japan}
\footnote{\href{mailto:jie.sheng@ipmu.jp}{jie.sheng@ipmu.jp} (Corresponding Author)}}
\author[1,2,4]{{\large Chuan-Yang Xing} \footnote{\href{mailto:chuan-yang-xing@sjtu.edu.cn}{chuan-yang-xing@sjtu.edu.cn}}
\affil[4]{
College of Science, China University of Petroleum (East China), Qingdao 266580, China}}
\date{}

\maketitle

\vspace{-2mm}
\begin{abstract}
\fontsize{10pt}{12pt}\selectfont

The dark matter scattering with atomic bound electrons
is a crucial avenue for exploring the sub-GeV mass range.
The commonly used factorization, where atomic effects
are encoded in an overall form factor multiplying the
free-electron scattering matrix element, is not necessarily true.
Especially, the free-electron kinematics and phase space cannot
consistently apply for off-shell bound electrons.
Starting from the first principles of quantum field theory,
we establish a theoretically consistent formalism
to account for the atomic effects.
By taking the scalar-type interaction as an example,
we investigate the difference between the non-relativistic
and relativistic calculations to show that the
relativistic effects can lead to a $30\% \sim 50\%$
reduction in the scattering phase space and differential cross section. In other words,
not just a theoretically consistent formalism for
the atomic effects but also relativistic calculation
with Dirac equation are necessary.

\end{abstract}

\section{Introduction}

More than $80\%$ of the matter in our Universe today is dark matter (DM) \cite{Bertone:2004pz,Young:2016ala,Arbey:2021gdg,Cirelli:2024ssz}.
However, the particle nature of DM and its interaction with the standard model (SM) particles are
still unknown. One of the experimental strategies to search for DM is the direct detection based 
on DM-nucleus scattering \cite{ Goodman:1984dc,MarrodanUndagoitia:2015veg,Billard:2021uyg,Akerib:2022ort}.
For non-relativistic DM in our galaxy, only those with mass
$m_\chi \gtrsim \mathcal{O}(1)\,$GeV can make the nucleus
recoil to overcome the detection threshold efficiently. 
The DM with a mass of the GeV to TeV range, 
which corresponds to the Weakly Interacting Massive
Particle (WIMP) \cite{Jungman:1995df,Arcadi:2017kky,Roszkowski:2017nbc} DM, 
has already received strong constraints from direct detection
experiments \cite{Schumann:2019eaa,Akerib:2022ort
}, such as PandaX \cite{PandaX:2018wtu,PandaX-4T:2021bab},
LZ \cite{LZ:2011rhn,LZ:2022lsv}, and Xenon \cite{XENON:2023cxc,XENON:2024wpa}
that have reached tonne scale. On the other hand,
light DM in the sub-GeV mass range still has large
parameter space \cite{Essig:2022dfa} and becomes more
popular nowadays \cite{Cooley:2022ufh}.

To ensure efficient energy transfer from sub-GeV DM particles to overcome the kinematic threshold in direct detection experiments, the scattering target should have a comparable mass within the sub-GeV mass range \cite{goldstein:mechanics,Kahn:2021ttr}.
In this context, the electrons in existing direct detection experiments serve as promising targets \cite{Essig:2011nj}. However, electrons of a usual direct detection experiment are always in bound quantum states within the Coulomb potential of nucleus.
Comparing with the free electron state, 
there are several non-negligible differences. Firstly, bound electrons have a negative binding energy, leading to a modification of their particle dispersion relationship. Secondly, the wave function of a bound electron is non-trivial, resulting in a distribution of incoming momentum rather than a fixed value. This, in turn, changes its recoil spectrum accordingly.
Therefore, it is crucial to establish a comprehensive theoretical description of DM-bound electron scattering
that incorporates all atomic effects.

Historically, the scattering with atomic electrons was initially
studied in the context of neutrino-electron interactions \cite{Gaponov:1977gr}.
In those pioneering works of 1990s, the relativistic and many-body effects in both initial and final electron wave functions were accounted for \cite{Fayans:1992kk,Dobretsov:1992qh,Fayans:2000ns,Kopeikin:2003bx}, and atomic form factors in the non-relativistic limit were analytically derived \cite{Morgunov:1996jk}.
Focusing exclusively on the light-nuclei
atomic scattering, the subsequent researches took a
step back to consider only the initial-state wave
functions and binding energies of bound electrons
\cite{Gounaris:2001jp,Gounaris:2004ji}. Later, the summation over all possible quantum states of the final electron and their corresponding wave functions were revisited in \cite{Voloshin:2010vm,Wong:2010pb}. Within the extensively studied low-energy scattering framework, the electron wave functions admit solutions under the non-relativistic approximation \cite{Morgunov:1996jk,Kouzakov:2010tx,Kouzakov:2011vx,Kouzakov:2011ka,Kouzakov:2014pua,Kouzakov:2014yha}. More recent works have further explored the influence of relativistic
\cite{Fayans:1992kk}
and many-body
\cite{Fayans:1992kk,Morgunov:1996jk,Chen:2013lba,Chen:2016lyr}
effects
on wave functions and scattering cross sections \cite{Chen:2013lba,Chen:2014dsa,Chen:2014ypv,Chen:2016lyr}.

Since both neutrino and DM particles are electrically
neutral, the same calculation may apply for both of them \cite{Chen:2013iud,Chen:2016lyr,Wu:2017zcd}. The research on atomic effects in the field of DM detection follows a similar pattern. 
Initially, the study of the Inverse Primakoff process of axions on atomic electron targets used a crude approximation of the electron Gaussian distribution \cite{PhysRevD.37.618}.
Of the first several attempts for the atomic effects
of DM-electron scattering, \cite{Bernabei:2007gr}
uses the momentum distribution function while
\cite{Dedes:2009bk} uses the wave function in the
coordinate space, both for the initial bound electron.
The other paper \cite{Kopp:2009et} takes both the
initial and final electron wave functions into
consideration, only the excited state is implemented
for the final-state electron. The ``{\it unbound
wavefunctions}'' are taken into account first in
\cite{Essig:2011nj}.
Later, the atomic effects are 
included as a universal form factor together with the $\mathcal{M}$-matrix of free-electron scattering \cite{Essig:2012yx,Chen:2015pha,Essig:2015cda,Roberts:2015lga,Roberts:2016xfw,Roberts:2016sem,Essig:2017kqs,Bloch:2020uzh,Chen:2021qao,Hamaide:2021hlp,Ge:2021snv,Emken:2021vmf}. 
Within this framework, the final-state electron wave functions have been incorporated, where the scalar atomic form factor is defined as the inner product of the initial- and final-state wave functions.
The influence of final-state interactions \cite{Plestid:2024jqm} 
has also be taken into account.
Recently, the influence of different DM-electron interactions on the form of the atomic form factor has been studied \cite{Catena:2019gfa,Liu:2021avx,Ge:2021snv,Wu:2022jln,Liang:2024lkk,Liang:2024ecw,Krnjaic:2024bdd,Liu:2024bhy}.
Under the non-relativistic  effective field theory, they are reduced to the above-mentioned scalar-type form factor.

The calculation of atomic factor depends on the specific form of wave functions \cite{Chen:2013iud,Chen:2013lba,Chen:2014ypv,Chen:2016eab,Wu:2017zcd,Pandey:2018esq,Roberts:2019chv,Qiao:2020ybv,Dent:2020jhf,Ge:2022ius,Wu:2022psn,Gao:2020wer,Abe:2020mcs}.
In the non-relativistic limit, they are
simply the solutions of the Schrödinger equation usually used in literature \cite{Essig:2011nj,Essig:2012yx,Essig:2015cda,Chen:2015pha,Essig:2017kqs,Catena:2019gfa,Ge:2021snv}. Additionally, several packages, like \texttt{QEdark} \cite{Essig:2015cda,softwareYu} and \texttt{DarkARC} \cite{Catena:2019gfa,2021ascl.soft12011E}, have been developed to compute non-relativistic atomic form factors.
However, the relativistic effects are naturally required.
For an electron inside an atom, its kinetic energy can be estimated from the Virial theorem as $T_e \simeq \alpha^2 Z_{\rm{eff}}^2 m_e/2$, with $\alpha$ being the fine structure constant and $Z_{\rm{eff}}$ being the effective nuclear charge. Taking the electron in the innermost layer of a Xenon atom for illustration, 
it has an effective charge $Z_{\rm{eff}} \sim 50$ 
and a kinetic energy $T_r \sim 0.1 m_e$,
which has already approached the relativistic regime
since the corresponding electron velocity $\sqrt{2 T_r / m_e}$
is already around 40\% of the speed of light.
Furthermore, in certain specific processes, such as the boosted DM scattering \cite{Yin:2018yjn, Cappiello:2018hsu, Bringmann:2018cvk,Ema:2018bih} and DM absorption \cite{Pospelov:2008jk,An:2014twa,Bloch:2016sjj,Dror:2019onn,Dror:2019dib,Dror:2020czw},
the energy transfer can also be of 
$\mathcal{O}(0.1) m_e$.

To study the relativistic effects, the wave functions
should be solved from the covariant Dirac equation
\cite{Roberts:2015lga,Roberts:2016xfw,Bloch:2020uzh,Whittingham:2022wxc}.
Still assuming the factorization formalism with
a universal form factor \cite{Roberts:2015lga,Roberts:2016xfw,Roberts:2016sem,Bloch:2020uzh},
%These papers applied the relativistic wave function using the direct factorization formalism: 
%[52]: eq.2 
%[53]: eq.3
%[54]: it is the phd thesis with same formalism  
%[56]: eq.6.6 comparison of NR-to-R, fig.15 right panel, almost the same.
it was initially found that using the approximated
relativistic wave functions significantly increases
the atomic form factor \cite{Roberts:2015lga,Roberts:2016xfw,Roberts:2016sem}.
%[52-54]: This group concluded an huge enhancement due to relativistic effects, [52]fig.2 [53] fig.4 
% [52], Eq.(15)
The relativistic effects were later
explored through the wave functions 
solved via the random phase approximation \cite{Chen:2016eab}
% [69]: This paper concluded a suppression due to relativistic effects. It is explicitly written in abstract:  " It is found that the
%atomic binding effect yields a sizable suppression to the neutrino-electron scattering cross section at low
%recoil energies. Compared with the previous calculation based on the free electron picture, our calculated
%event rate of electronic recoil in the same detector configuration is reduced by about 25%" Their calculation method is the "we adopt an ab
%initio many-body method: the relativistic random phase approximation (RRPA) [13–15]" written in introduciton
and the
full Dirac equation \cite{Whittingham:2022wxc},
%[90]: Applying QFT method (similar to us) to nu-e scattering, concluded a suppression due to dirac wave function: eq.27+fig1 and fig2
with reduced scattering cross section. More recent analyses \cite{Alhazmi:2025nvt}
%[91]: Different interaction gives different results of comparison: fig3 + fig4
further indicate that whether this corrections lead to an enhancement or reduction depends on the interaction type and the kinematic regime. However, these effects remain inconclusive and the underlying physical origin has not been fully clarified.

In order to comprehensively study the atomic effects and maintain theoretical consistency,
the bound-electron scattering 
should be established within 
the framework of quantum field theory (QFT)
\cite{Ge:2021snv,Whittingham:2021mdw,Whittingham:2022wxc}
with electrons being not free states but affected by the
atomic Coulomb potential. 
This approach has been previously applied to the DM scattering
with a non-relativistic electron \cite{Ge:2021snv} as well as
the neutrino–electron scattering \cite{Whittingham:2021mdw,Whittingham:2022wxc}.
In this paper, we further develop such QFT formalism for the
DM-bound electron scattering cross section from the
first principles. Especially, we use the second quantization to describe the
atomic electrons as bound or ionized states \cite{Ge:2021snv},
instead of using the free-electron fields
\cite{Essig:2015cda} or treating atomic electrons only as wave functions
\cite{Roberts:2015lga,Roberts:2016xfw,Roberts:2019chv}.
This formalism is extendable and universally
applicable in the relativistic regime. 
For the first time, we provide a detailed analysis to
explain the physical origin of the relativistic effects.
Particularly, the relativistic corrections arise from not
only the electron kinetic energy but also the phase shifts
of the ionized wave functions.

The paper is organized as follows. 
In \gsec{sec:Factorization}, we review the previous 
literature in more detail and demonstrate the 
necessity of our new formalism within the 
framework of QFT. Especially, the factorization
with free electron scattering amplitude would
lead to negative cross section in some situations.
Then, we discuss 
the second quantization of electron fields
in the atomic Coulomb potential
and derive the
cross section of bound electron-DM scattering in \gsec{sec:scatteringFormalism}. 
In \gsec{sec:factorization}, we factorize the scattering matrix element and review the atomic factor in non-relativistic limit.
After that, we make a comparison between the relativistic
and non-relativistic atomic factors to investigate the
relativistic effects in \gsec{sec:relativisticEffects}. Finally, 
we give our conclusion in \gsec{sec:conclusion}.

\section{Factorization with Free Electron Scattering Amplitude}
\label{sec:Factorization}

In the usual QFT treatment of scattering, the external
particles are free and hence can be described by plane
waves with definite momentum if the interaction
is not strong enough to affect the asymptotic states which
is known as the Born approximation.
It is then of great
convenience to derive the scattering amplitude and
cross section in the momentum space
\cite{Peskin:1995ev,Weinberg:1995mt}.
However, in the DM direct detection experiments,
the initial-state electrons are typically bound inside
atoms. Although the final-state
recoil electron can leave atom as ionization to be
detected, it is still subject to the nuclear charge.
Currently, the first
ton-scale DM direct detection experiments (PandaX-4T
\cite{PandaX:2018wtu,PandaX-4T:2021bab},
LZ \cite{LZ:2011rhn,LZ:2022lsv}, and
XENONnT \cite{XENON:2023cxc,XENON:2024wpa})
are all Xenon-based detectors.
The nuclear charge can be as high as $Z = 54$.
For both the initial and final states, the Coulomb
attractive potential from the nuclear charge distorts
the electron wave functions significantly
from plane waves \cite{Sakurai:2011zz}.

Such wave function distortion should have significant
effect on the scattering process and hence the
DM detection prospects. First, with the initial-state
electron possessing atomic ``{\it fermi motion}'', the recoil
electron spectrum takes different distribution.
As demonstrated in our previous study
on the fermionic DM absorption on electron targets,
the originally mono-energetic spectrum with free electron
target broadens to a characteristic peak
\cite{Ge:2022ius}. Second, the atomic effect reduces
the scattering cross section in the low-energy region
\cite{Ge:2021snv,Ge:2022ius}. These atomic effects
with bound electron target and ionized recoil final
state need to be taken into account.

The existing treatment of the atomic effects
is based on the assumption that the scattering matrix
element $|\mathcal M|^2$ can be simply decomposed as a
direct product of the scattering matrix element
$|\mathcal{M}_{\rm free}|^2$ with free electron and
an overall atomic form factor $|f|^2$
\cite{Essig:2011nj,Essig:2012yx,Essig:2015cda},
\begin{equation}
|\mathcal{M}|^2
\equiv
\left|\mathcal{M}_{\text {free }}\right|^2 \times \left|f_{i \rightarrow f}({\bm q})\right|^2,
\quad
{\rm where}
\quad
f_{i \rightarrow f}({\bm q})
\equiv
\int \frac{d^3 {\bm k}}{(2 \pi)^3}
\phi_f^*({\bm k}+{\bm q}) \phi_i({\bm k}).
\label{previous_f}
\end{equation}
The atomic form factor is a function of the initial
bound state $\phi_i ({\bm k})$ and final ionized
state $\phi_f ({\bm k})$ wave functions in the momentum
space.

However, this simple decomposition is not fully self-consistent 
and can sometimes introduce issues with 
negative $|\mathcal M|^2$ or equivalently
negative differential cross section.
Taking 
the vector interaction mediated through a 
dark photon,
$
\mathcal{L} 
\supset 
g_\chi \bar{\chi} \gamma^\mu \chi A_\mu^{\prime}
+
g_e \bar{e} \gamma^\mu e A_\mu^{\prime}+\frac{1}{2} m_{A^{\prime}}^2 A_\mu^{\prime} A^{\prime \mu}$,
for illustration, the corresponding free matrix
element is \cite{Cao:2020bwd},
\begin{align}
    \overline{|\mathcal{M_{\text{free}}}|^2}
=
    \frac{32 m_\chi 
    [m_e T_r^2 - T_r
    (m_e^2 + m_\chi^2 + 2 m_e m_\chi + 2 m_e T_\chi) 
    +
    2 m_e (m_\chi + T_\chi)^2
    ]}{(t -m_{A'}^2)^2},
\label{Mdp}
\end{align}
where $T_\chi$ represents the initial DM kinetic 
energy and $T_r$ is the final-state electron
recoil energy. For generality, we keep the electron
($m_e$), dark matter ($m_\chi$), and dark photon
($m_{A'}$) masses without assuming hierarchical
structure among them.
Below we will try to illustrate how
$|\mathcal M_{\text{free}}|^2$ develops negative values in the
allowed parameter space.

Being free particles, the initial- and final-state
DM particles follow the usual on-shell conditions, 
\begin{gather}
  E_{\chi}
=
 \sqrt{ m^2_{\chi} + 
 |{\bm p}_\chi|^2  }
\,,
\quad
    E'_{\chi}
=
 \sqrt{ m^2_{\chi} + 
 |{\bm p}_\chi|^2  + |{\bm q}|^2 - 2 |{\bm p}_\chi| |{\bm q}| \cos \theta_q },
 \label{eq:Echi}
\end{gather}
where ${\bm p}_\chi$ is the initial DM momentum and
${\bm q}$ is the momentum transfer. Using the energy
conservation law, 
\begin{equation}
    E'_{_\chi}
=
    E_{\chi} - \Delta E\,,
\quad \text{with} \quad
    \Delta E 
\equiv
    |E_b| + T_r,
\label{energycon}
\end{equation}
the final-state DM energy $E'_\chi$ can be expressed
in terms of the incoming one $E_\chi$ and the energy
transfer $\Delta E$. The opening angle $\theta_q$
between $\bm p_\chi$ and $\bm q$ varies from 0
to $\pi$ and consequently $|\cos \theta_q| \leq 1$.
Then the momentum transfer $|\bm q|$ is constrained
to be in the range of, 
\begin{equation}
% |{\bf q}|_{\min} \equiv
  |{\bm p}_\chi|
- \sqrt{(E_{\chi} - \Delta E)^2 - m^2_{\chi}}
\leq
  |{\bm q}|
\leq
  |{\bm p}_\chi|
+ \sqrt{(E_\chi - \Delta E)^2 - m^2_{\chi}}.
% \equiv |{\bf q}|_{\max}.
\label{eq:q-limits}
\end{equation}

The energy transfer $\Delta E$ is summation of the 
binding energy $E_b$ of the initial electron and the
final-state electron recoil energy $T_r$. 
The initial DM kinetic energy $T_\chi$ must be
larger than the electron binding energy $|E_b|$,
$T_\chi > |E_b|$, to ionize it.
Since the binding energy is negative, we use its explicit
absolute value $|E_b|$ to avoid ambiguity. A direct
consequence is that $\Delta E > 0$.
The quantity inside the square root of \geqn{eq:q-limits}
should be positive,
$\Delta E < E_{\chi} - m_\chi \equiv T_\chi$
or $\Delta E > E_{\chi} + m_\chi $.
The latter solution is excluded because
the final DM energy  
$E'_\chi$ should be positive and 
so that $\Delta E < E_\chi$
according to \geqn{energycon}. 
Consequently, the electron recoil energy 
$T_r = \Delta E - |E_b| < T_\chi - |E_b|$ defined in \geqn{energycon} has a range of $T_r \subset (0, T_\chi - |E_b|)$.

\begin{figure}[t]
\centering
\includegraphics[width=0.68\textwidth]{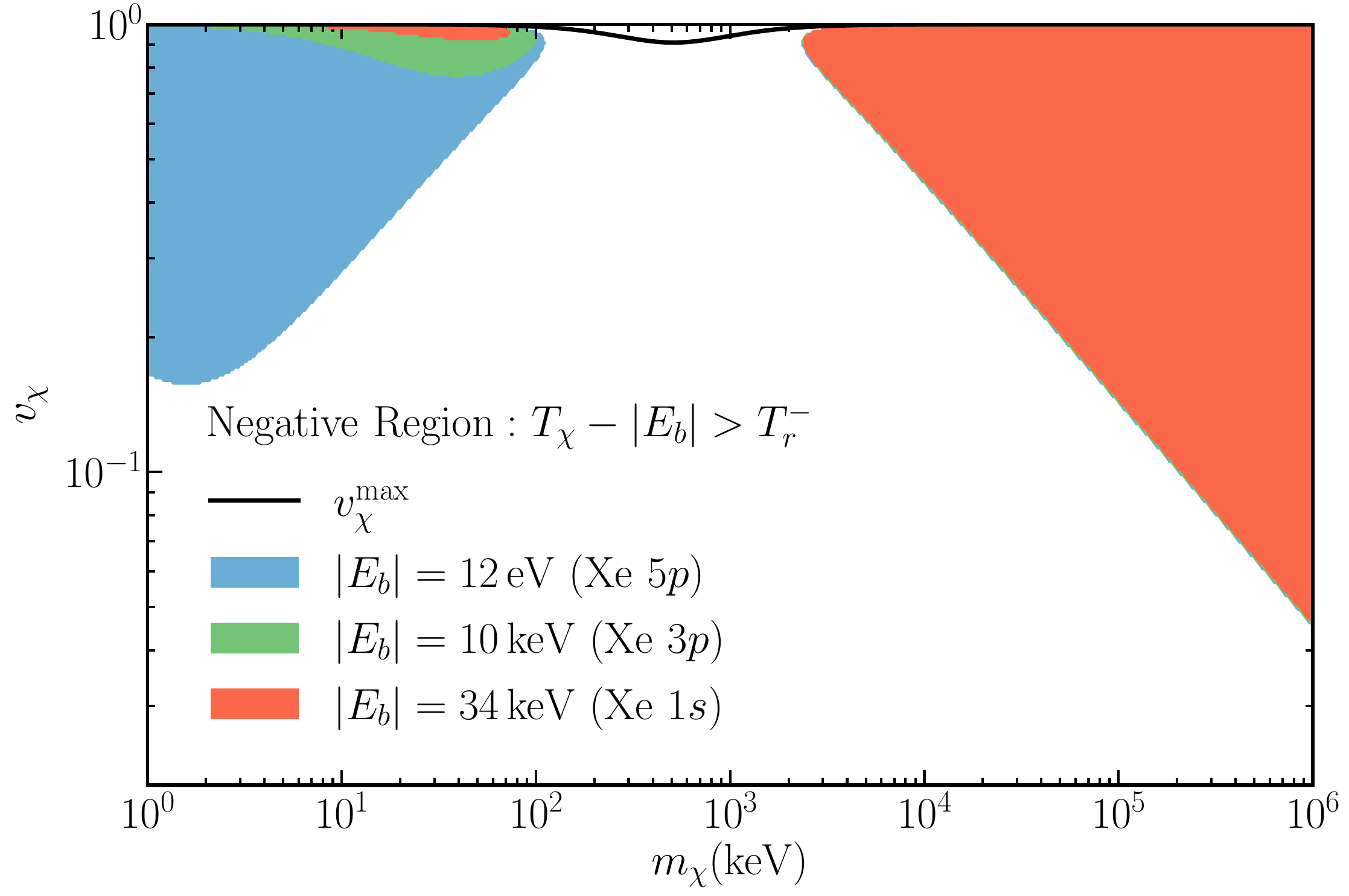}
\caption{
The colored parameter space in which the scattering
matrix element with free electron becomes negative 
with the bound-electron kinematics.
The (blue, green, red) region
corresponds to the bound electron with a binding
energy of ($0.012$, $10$, $34$)\,keV.
The black solid line corresponds to the maximum DM
velocity that allows the differential
cross section to potentially take negative values.
}
\label{nagativeRegion}
\end{figure}

The free matrix element \geqn{Mdp} is a quadratic
function of $T_r$. Once the recoil energy $T_r$ lies
in the range, $T^-_r < T_r < T^+_r$ with
\begin{align}
  T^\pm_r
\equiv
  \frac 1 {2 m_e}
\left[
  (m_e + m_\chi)^2
+ 2 m_e T_\chi
\pm
  \sqrt{(m_e^2 + m_\chi^2 + 2 m_e m_\chi + 2 m_e T_\chi)^2 - 8 m_e^2 E^2_\chi}
\right],
\end{align}
the differential cross section would become negative.
This happens when
\begin{align}
  T_\chi
<
  \frac {(m_e + m_\chi)^2 - 2 \sqrt 2 m_e m_\chi}
        {2 (\sqrt 2 - 1) m_e},
% \quad \Rightarrow \quad
%   v_\chi
% <
%   \frac {(m_e + m_\chi)^2 - 2 \sqrt 2 m_e m_\chi}
%        {(\sqrt 2 - 1) m_e m_\chi}.
\end{align}
which means there is an upper limit on the DM velocity
for the negative value region shown as the black line
in \gfig{nagativeRegion}. It clearly shows that
there is a huge parameter space with the DM velocity limit
close to the speed of light that potentially allows
the differential cross section to assume negative values.

Once the above range overlaps with the physical range of recoil energy, 
$T_r \subset (0, T_\chi - |E_b|)$
or equivalently
$T^-_r < T_\chi - |E_b|$,
the matrix element \geqn{Mdp} will have negative values,
which gives $T_\chi^- < T_\chi < T_\chi^+$ where,
\begin{equation}
  T_\chi^{\pm}
\equiv 
  \frac{(m_e - m_\chi)^2}{2m_e} 
   \pm
  \frac{\sqrt{(m_e -m_\chi)^4 -8 m_e^2 m_\chi^2 - 4 m_e(m_e+m_\chi)^2 |E_b| - 4 m_e^2 |E_b|^2}}
  {2 m_e}.
  \label{Txpm}
\end{equation}
For illustration, the negative regions with the
$5p$ ($|E_b| = 12$\,eV), $3p$ ($|E_b| = 10$\,keV), and $1s$ ($|E_b| = 34$\,keV)
electron in the Xenon atom have been shown as blue,
green, and red in \gfig{nagativeRegion}, respectively.
To ensure the positivity of the expression under the square root in \geqn{Txpm}, 
the DM mass can not be too close to the electron mass.
For $m_\chi \ll m_e$, the leading terms are 
$m_e^4 - 4 (m_\chi + |E_b|) m_e^3$, and its positivity enforces $m_\chi  < m_e/4 - |E_b|$. 
For Xenon atom, the maximum electron binding energy is $|E_b| = 34\,$keV, the above condition roughly gives $m_\chi \lesssim 100\,$keV, and a smaller binding energy corresponds to a larger
DM mass boundary. 
For $m_\chi (m_e) \gg |E_b|$, the leading-order term $(m_e-m_\chi)^4 - 8m_e^2 m^2_\chi$
dominates, and its positivity condition requires $m_\chi > (1+\sqrt{2} + \sqrt{3+2\sqrt{2}}) m_e \simeq 5 m_e$. 
Consequently, regions in
\gfig{nagativeRegion} where the DM mass lies outside
these two intervals appear as blank zones which are
safe from the issue of negative cross section. 
Furthermore, when DM mass is sufficiently large, 
$E_b$ becomes negligible, leading to nearly identical negative-value regions predicted by different electron orbitals. In the low-mass regime, the colored region shrinks with increasing $E_b$, as only DM with significantly high velocities can attain the kinetic energy necessary for electron ionization.
The negative-value criterion also prefers relativistic DM with large
velocities as $v_\chi \gtrsim 10^{-2}$.
Generally, the velocity of halo DM in the laboratory frame is approximately 
$v_\chi \sim 10^{-3}$ and hence the problem does not arise.
Therefore, this negative-value issue becomes more pronounced
for accelerated DM, such as the Cosmic Ray-boosted DM (CRDM)
\cite{Bringmann:2018cvk,Ema:2018bih,Ge:2020yuf,Cao:2020bwd,Granelli:2022ysi,Xia:2024ryt,Guha:2024mjr,PandaX:2024pme}.

Taking the N-shell (principle quantum number $n=4$) electron
in the Xenon atom
with binding energy $|E_b| = 5\,$keV as an example. 
If the incoming DM has a mass $m_\chi = 10\,$keV and a kinetic energy $T_\chi = 10\,$keV, one can calculate that $T_r^- = 1.5\,$keV, $T_r^+ = 549.7\,$keV, and $T_\chi - |E_b| = 5\,$keV. 
The matrix element becomes negative in the overlapping 
range $T_r \subset (1.5, 5)\,$keV.
For instance, the numerator in 
\geqn{Mdp} is 
$-1.3\times 10^{-4}\,$MeV$^4$
if $T_r = 3\,$keV.

The negative $|\mathcal M|^2$ and hence negative differential
cross section arise from the inconsistency of imposing
the free-electron phase space and kinematics onto the
scattering with bound electron. Especially, a bound atomic
electron has fermion motion with initial momentum such that
for the same momentum transfer from the DM side the final-state
electron recoil can have much wider distribution than the
free-electron case.
Although the issue manifests only in a special region
of the DM parameter space with high DM velocity, it already
indicates a fundamental inconsistency in the theoretical
description. A well-defined framework should be applicable
across the entire kinematic parameter space.
Below, we detail a consistent
description of the DM scattering with bound electron
from first principles
with a brief summary as conclusion to be found
at the end of \gsec{sec:factorization}.

\section{DM-Bound Electron Scattering from First Principles}
\label{sec:scatteringFormalism}

In the usual QFT formalism, the scattering process starts from
the asymptotic state in the infinite past ($t \rightarrow - \infty$)
where the participating particles are sufficiently far apart such that 
they are not yet interacting. After scattering, the process
ends at the infinite future ($t \rightarrow + \infty$)
where the particles have become so far 
apart again that they have ceased interacting. 
For the free particle scattering, particles are not
subject to any interaction at infinite times ($t \rightarrow \pm \infty$) 
but the free Hamiltonian that gives the free particle solution
labeled with eigenvalues of momentum $p$ and spin $s$
\cite{Peskin:1995ev,Weinberg:1995mt}.

However, the electrons bound within atoms are no
longer free particles. The Born approximation
no longer applies for the atomic electrons.
Correspondingly, a bound electron
has different wave function and eigenvalue than a free electron.
For the bound electron state in a central field,
the good quantum number is its energy and the total
angular momentum rather than its momentum or spin.
Consequently, the DM scattering with a bound electron
needs a different formalism to derive the scattering
matrix element and cross section.

\subsection{QFT Formalism of Scattering with Bound Electron}

To obtain a theoretically self-consistent formalism,
we first reexamine and derive how scattering involves
bound electrons within the framework of QFT.
The calculation starts from the most fundamental
definition of cross section as an integration of
probability over the area spanned by the
two-dimensional impact parameter $\bm b$
\cite{Peskin:1995ev},
\begin{eqnarray}
  \sigma
\equiv
  \int P({\bm b}) d^2{\bm b}
\quad
\mathrm{where}
\quad
  P 
\equiv 
  |{}_{\text{out}}\braket{\phi_\chi \phi_e |\phi_\chi \phi_e  }_{\text{in}}|^2.
\label{eq:sigma}
\end{eqnarray}
The impact parameter ${\bm b}$ describes the 
transverse displacement of the incoming
wave packet and the probability $P$ 
is an inner product of the $in$
($|\phi_\chi \phi_e \rangle_{\text{in}}$) and $out$
(${}_{\text{out}}\langle \phi_\chi \phi_e|$)
states.

In the conventional scenario of free particle scattering, 
the asymptotic states are the eigenstates of the free
Hamiltonian with eigen-momentum $p$ and spin $s$,
namely, the plane waves.
However, the situation changes a lot when 
the electron scattering happens inside an atom
in the presence of Coulomb potential. 
Even at the infinite times $t\rightarrow \pm \infty$
that are considerably distant from the scattering time,
the electron can still be subject to the influence of
the atomic Coulomb potential. For convenience, 
the initial bound electron state is denoted as $\ket{i}$.
Then the $in$ state of the scattering system is
constructed as, 
\begin{equation}
|\phi_\chi \phi_e \rangle_{\text{in}}
\rightarrow
\int \frac{d^3 {\bm p}_\chi }{(2 \pi)^3} 
\frac{\phi_\chi
\left(\bm{p}_\chi \right) 
e^{-i \bm{b} 
\cdot \bm{p}_\chi }}{\sqrt{2 E_{{\bm p}_\chi}}}
\left|
i;
\bm{p}_\chi 
\right\rangle_{\text {in }},
\end{equation}
Being a free particle, the initial DM
takes a momentum eigenstate
$\ket{{\bm p}_\chi}$. Without knowing the DM
momentum, the initial DM state
$\ket{\phi_\chi}_{\text{in}}$ 
should in general be defined as an integration
over the wave function $\phi_\chi ({\bm p_\chi})$ \cite{Peskin:1995ev}. 
The normalization condition of the wave function,
$ \int d^3 {\bm p}_\chi
  |\phi_\chi({\bm p}_\chi)|^2 / (2 \pi)^3 = 1$,
requires the wave function dimension to be $[\phi_\chi({\bm p_\chi})] = -3/2$.
For the final-state particles, the $out$ state
contains the asymptotic momentum eigenstate of DM,
$\ket{{\bm p}'_\chi}$, and the energy eigenstate
of the final electron $\ket f$ as,
\begin{equation}
    |\phi_\chi \phi_e \rangle_{\text{out}}
\rightarrow
    \ket{{\bm p'}_\chi ;f }.
\label{eq:final}
\end{equation}
While the initial electron is in a bound state,
the final electron should take the ionized state.
The final-state
electron can also take the excited state without
leaving the atom and the detection can be achieved
with the de-excitation photon.
For illustration, we consider only the
ionized state for the final-state electron.

The state evolution from the infinite past to the infinite
future is described by the Hamiltonian $H_{\rm int}$.
In th interaction
picture, the so-called $\mathcal S$ matrix can be defined
as the unitary operator $S \equiv e^{-i H_{\rm int} t}$
sandwiched by the asymptotic states \cite{Peskin:1995ev}, 
\begin{equation}
\mathcal{S}
\equiv
{ }_{\text {out}}\left\langle
\phi_e \phi_\chi \mid 
\phi_e
\phi_\chi \right\rangle_{\text {in }} \equiv
\left\langle
f; \bm{p}'_\chi \mid 
S \mid
i;
\bm{p}_\chi \right\rangle.
\label{S_Free}
\end{equation}
Subsequently, the transition matrix element $\mathcal T$ can be 
extracted from the decomposition $S \equiv {\bf 1 } + i T$
of the $S$ operator, 
\begin{equation}
  \mathcal T
\equiv
    \braket{f;{\bm p}_\chi'|
    i T
    |i ;{\bm p}_\chi}.
\label{Tmatrix}
\end{equation}
By applying the aforementioned definitions,
the cross section now involves both 
the integration over the impact 
parameter ${\bm b}$ and final state
phase space as,
\begin{equation}
  \sigma
\equiv 
    \int d^2{\bm b}
    \int d \Omega'_e
  \dFs{{\bm p}_{\chi}'}
  \left|
  \int \dFp{{\bm p}_\chi}
  e^{i {\bm p}_\chi\cdot {\bm b}}
  \phi_{\chi}({\bm p}_\chi)\,\,
  \mathcal{T} ({\bm q})
  \right|^2,
\label{cs1}
\end{equation}
where $\bm q \equiv \bm p_\chi - \bm p'_\chi$ is the
momentum transfer. The phase space integration
for the initial DM is at the field or amplitude
level while the one for the final DM is at the
cross section level \cite{Peskin:1995ev}.

To follow the convention of free-particle scattering, 
it is convenient to decompose the $\mathcal T$ matrix
into a $\mathcal{M}$ matrix and some $\delta$ functions.
Since the atomic electrons are eigenstates of
just energy but not momentum, only the $\delta$ function
for energy conservation can be extracted,
\begin{equation}
    \mathcal{T} ({\bm q})
\equiv 
    \mathcal{M} ({\bm q})  
    (2 \pi) \delta \left( \Delta E - E_{{\bm p}_\chi} + E_{{\bm p}'_\chi} \right),
\label{decoupleT}
\end{equation}
where $E_{\bm p_\chi}$ ($E_{\bm p'_\chi}$) is the
initial (final) DM energy and
$\Delta E \equiv E'_e - E_e$ is the electron energy difference.
To ensure the correct dimension of the two-body scattering cross section, $[\sigma] = -2$, the dimensions of the 
$\mathcal T$($\mathcal{M}$) matrix and the
electron final-state phase space should
satisfy the relation
$[\mathcal{T}] = [\mathcal{M}] - 1 = - [d \Omega'_e]/2 - 2$
whose concrete value needs to be determined later.
Only in this way, the transition probability
$P(\bm b)$ that corresponds to the second integral
of \geqn{cs1} is dimensionless.

Expanding the square explicitly and 
redefining a dual set of momentum $\bar{\bm p}_\chi$ for
the incoming DM, the cross section in \geqn{eq:sigma}
becomes, 
\begin{align}
    \sigma
& = 
    \int d \Omega'_e
  \int \dFs{{\bm p}_{\chi}'}
  \int d^2{\bm b}
  e^{i {\bm b}\cdot ({\bm p}_\chi - \overline{{\bm p}}_\chi)}
\nonumber
\\ 
& \times 
  \Biggl[
  \int \dFp{\overline{{\bm p}}_\chi}
  \phi_\chi(\overline{{\bm p}}_\chi)
  \mathcal{M}(\overline{{\bm p}}_{\chi},{\bm p}_{\chi}')
  (2\pi) \delta(E_{{\bm p}_{\chi}'} - E_{\overline{{\bm p}}_{\chi}} + \Delta E)
  \Biggr]^*
\nonumber
\\ 
& \times 
  \Biggl[
  \int \dFp{{\bm p}_\chi}
  \phi_\chi({\bm p}_\chi)
  \mathcal{M}({\bm p}_{\chi},{\bm p}_{\chi}')
  (2\pi) \delta(E_{{\bm p}_{\chi}'} - E_{{\bm p}_{\chi}} + \Delta E)
  \Biggr].
\end{align}
The integration of the impact parameter ${\bm b}$ results
in the momentum conservation perpendicular to it,
$(2\pi)^2\delta^{(2)}({\bm p}^{\perp}_\chi-\overline{{\bm p}}_{\chi}^{\perp})$.
Together with one of the two energy conservation $\delta$
functions, one can also eliminate the integration over
the parallel momentum ${\overline{\bm p}}^{\parallel}$ as,
\begin{eqnarray}
\left[
  (2\pi)^2\delta^{(2)}({\bm p}^{\perp}_\chi - \overline{{\bm p}}_{\chi}^{\perp})
  d^2 \overline{\bm p}_{\chi}^{\perp}
\right]
\left[
  (2 \pi) \delta(E_{{\bm p}_{\chi}'} - E_{\overline{\bm p}_{\chi}} + \Delta E)
  d \overline{\bm p}_{\chi}^{\parallel}
\right]
=
  \frac{(2 \pi)^3}{|\overline{{\bm p}}_{\chi}^{\parallel}|/E_{\overline{{\bm p}}_{\chi}}}
=
  \frac{(2 \pi)^3}{|{\bm v}_\chi|},
\end{eqnarray}
The integration over $d^3 \overline{{\bm p}}$ gives
the DM flux factor, $(2 \pi)^3/|{\bm v}_\chi|$.
The initial electron bound state does not have a
definite momentum and for the atom
as a whole the momentum is zero.
As a result, the incident velocity of DM essentially
represents the relative velocity between DM and
atom in the laboratory frame.

The above integration $d^2 \overline{\bm p}^\perp_\chi$
matches the dual momentums, $\overline{\bm p}^\perp_\chi = \bm p^\perp_\chi$.
Based on this, the two $\delta$ functions for energy
conservation essentially requires
$\overline{\bm p}^\parallel_\chi = \bm p^\parallel_\chi$.
In other words, the dual momentum $\overline{\bm p}_\chi$
is exactly its counterpart $\bm p_\chi$. Then the
other integration, $d^3 {\bm p}$, can be eliminated
by the normalization condition of the incoming DM
wave packet, $ \int d^3 {\bm p}_\chi
  |\phi_\chi({\bm p}_\chi)|^2 / (2 \pi)^3 = 1$.
After removing the dual momentum and the intial
DM wave function, the total cross section for the
ionization of a bound electron takes the form as, 
\begin{equation}
\sigma^{\rm ion}_{\rm bound} 
= 
    \frac{1}{2 |{\bm v}_\chi| E_{{\bm p}_{\chi}}}
    \int d \Omega'_e
  \int \dFs{{\bm p}_{\chi}'} 
  \overline{|\mathcal{M}({\bm p}_{\chi},{\bm p}_{\chi}')|^2}
  (2\pi) \delta(E_{{\bm p}_{\chi}'} - E_{{\bm p}_{\chi}} + \Delta E).
\label{CSnkm}
\end{equation}

Until now, we have established the general formalism for
the DM scattering with atomic electrons without making
assumption or simplification. Comparing with the
usual cross section formula for the scattering of
free particles, one major difference is the
final-state phase space integration,
\begin{align}
  \int d \Omega'_e
  \int \dFs{{\bm p}_{\chi}'} 
  (2 \pi)
  \delta(E_{{\bm p}_{\chi}'} - E_{{\bm p}_{\chi}} + \Delta E).
\label{eq:PS}
\end{align}
With the final-state electron in an ionized state,
its phase space $d \Omega'_e$ should also be
different from the free particle case. We will
try to establish the exact form of $d \Omega'_e$
according to the second quantization of the
electron ionized state as elaborated below.

\subsection{Second Quantization of Bound and Ionized States}
\label{subsec:2ndQuantization}

In the presence of Coulomb potential, an atomic electron
should follow the eigenstate $\psi_\alpha$ 
of the total Hamiltonian $H_B \equiv H_0 + H_{\rm int} + V$
with the free Hamiltonian $H_0$, the interaction
term $H_{\rm int}$,
and the Coulomb potential $V$.
While the interaction part $H_{\rm int}$
gives the DM-electron scattering, the electron
asymptotic states are subject to the Coulomb
potential $V$. Then,
the second quantization of electron should be different
from the free-particle case. Since the atomic electron
resumes an eigenstate of energy but not momentum, the
corresponding electron field can be generally defined as,
$e (x)
\equiv
  \sum_\alpha  
   a_\alpha \psi_\alpha (\bm{x}) e^{- i E_\alpha t}
$
with $\alpha$ generally denoting the
corresponding quantum numbers.
The anti-particle part in the field operator will be
omitted from now on for simplicity, since there is
no bound anti-electron in
the situation under consideration.
In atomic systems, the total energy of an electron
is typically very close to the electron mass.
For convenience, one may remove the electron mass
$m_e$ from its energy eigenvalue.
Since the atomic
Coulomb potential is independent of time, the time
component of the wave function is a phase factor 
$e^{- i E_\alpha t}$.
The space component $\psi_\alpha ({\bm x})$ are
the corresponding spatial wave functions of state
$\ket \alpha \equiv a^\dagger_\alpha |\Omega_0\rangle$
that is created by the operator
$a^\dagger_\alpha$ from vacuum $|\Omega_0\rangle$. 
This different quantization and scattering formalism based on the bound states of particles is the so-called
{\it Furry Picture} or {\it Bound-Interaction Picture}
\cite{Moortgat-Pick:2009fyg,Whittingham:2021mdw,Whittingham:2022wxc}.

The Coulomb potential energy of an electron 
is negative
to form a stable two- or many-body system.
Therefore, depending on its kinetic energy, the eigen-energy
of an electron can be negative, indicating a bound state, 
or positive, indicating an ionized state.
We will discuss the quantization of 
the bound and ionized states in \gsec{sec:bound}
and \gsec{sec:ionized} separately before deriving
the scattering cross section in \gsec{sec:xsec}.

\subsubsection{Bound State}
\label{sec:bound}

The atomic Coulomb potential is an isotropic central field.
In such a case, the angular momentum conservation law ensures 
the good quantum numbers of electrons 
being the total angular momentum $\kappa$ and the
magnetic quantum number $m_j$ of the total angular momentum $j$
\cite{cohen1977quantum}. 
In addition, bound states exhibit discrete energy levels. 
The remaining radial eigen-equation introduces
the principle quantum
number $n$ to denote the energy level. 
The field operator of a bound electron is composed of
the wave function $\psi_{n \kappa m_j}$ and the corresponding
annihilation operator $a_{n \kappa m_j}$,
\begin{equation}
  e_B(x)
\equiv
  \sum_{n \kappa m_j} a_{n \kappa m_j} 
  e^{-i E_{n \kappa} t} 
  \psi_{n \kappa m_j}(\bm{x}).
\label{bound_field}
\end{equation}

Following the convention in quantum mechanics,
the wave function normalization condition takes
the form as,
\begin{equation}
  \int \psi_{n \kappa m_j}^{\dagger}(\bm{x}) 
  \psi_{n^{\prime} \kappa^{\prime} m_j^{\prime}}(\bm{x}) 
  d^3 \bm{x}
=
  \delta_{n n^{\prime}} 
  \delta_{\kappa \kappa^{\prime}} 
  \delta_{m_j m_j^{\prime}},
\label{eq:bound-norm}
\end{equation}
which determines the wave
function dimension to be
$[\psi_{n \kappa m_j} ({\bm x})] = 3/2$. 
To ensure that the field operator has the same
dimension $[e_B ({\bm x})] = 3/ 2$ as the free-particle
case, the creation and annihilation operators
should be dimensionless,
$[a_{n \kappa m_j}] = [a^\dagger_{n \kappa m_j}] = 0$.
As a result, their commutation relations become,
\begin{equation}
\left\{
  a_{n \kappa m_j}, a_{n' \kappa' m'_j}^{\dagger}
\right\}
= 
\delta_{n n'}
\delta_{\kappa \kappa'}
\delta_{m_j m'_j},
\quad
\left\{a_{n \kappa m_j}, 
a_{n^{\prime} \kappa^{\prime} m_j^{\prime}}\right\}
= 
0,
\quad
\left\{a^\dagger_{n \kappa m_j}, 
a^\dagger_{n^{\prime} \kappa^{\prime} m_j^{\prime}}\right\}
= 
0,
\end{equation}
which differ from the case of a free particle. 

The vacuum state $\ket{\Omega_0}$ can be annihilated by 
the annihilation operator, $a_{n\kappa m_j} \ket{\Omega_0} = 0$, and its normalization remains to be $\braket{\Omega_0|
\Omega_0} = 1$.
Once the vacuum is defined, the bound states $\ket{n \kappa m_j}$ can be defined by a creation operator acting 
on vacuum as $\ket{n \kappa m_j} \equiv a^\dagger_{n \kappa m_j} \ket{\Omega_0}$. 
Under this definition, the inner product, or normalization condition, 
of electron states is, 
%be Lorentz invariant. For bound states, the Lorentz invariance is already satisfied as, 
\begin{equation}
    \braket{n' \kappa' m'_j | n \kappa m_j}
=
    \bra{\Omega_0} 
    a_{n' \kappa' m'_j} a_{n \kappa m_j}^\dagger 
    \ket{\Omega_0}
=
    \bra{\Omega_0} 
    \{a_{n' \kappa' m'_j}, a_{n \kappa m_j}^\dagger \}
    \ket{\Omega_0}
=
\delta_{nn'} \delta_{\kappa \kappa'}
    \delta_{m_j m'_j}.
\end{equation}
The wave function can be extracted by applying the
field operator to the particle state as,
$e_B(x) \ket{n \kappa m_j} =
e^{-i E_{n \kappa} t}\psi_{n \kappa m_j}(\bm{x}) \ket{\Omega_0}$.

\subsubsection{Ionized State}
\label{sec:ionized}

The ionized states share the same angular eigen-equations and 
good quantum numbers, $\kappa$ and $m_j$, with the bound case.
However, its energy is no longer discrete. 
As a result, the other good quantum number becomes the
kinetic energy
$T_r \equiv E - m_e = \sqrt{m_e^2 + |{\bm p}|^2} - m_e$
where the second equality holds in the asymptotic limit.
Note that $T_r$ is actually the electron recoil energy
that can deposit in a direct detection experiment.
In its radial component, the asymptotic state of the
ionized electron is a free state with momentum $|{\bm p}|$.
To incorporate a continuous asymptotic momentum,
the commutation of the annihilation and creation 
operator should become \cite{Marecki:2003db}
\begin{equation}
  \{ a_{T_r \kappa m_j}, a^\dagger_{T'_r \kappa'm'_j}\} 
=
  2 \pi \delta_{\kappa \kappa'} 
  \delta_{m_j m'_j} 
  \delta(T_r - T'_r)
\,,
\quad
\{ a_{T_r \kappa m_j}, a_{T'_r \kappa'm'_j}\} 
=
\{ a^\dagger_{T_r \kappa m_j}, a^\dagger_{T'_r \kappa'm'_j}\}
=
0.
\end{equation}
Now, the creation and annihilation operators have
mass dimension $[a_{T_r \kappa' m'_j}] = -1/2$. 

Although the creation and annihilation operators are
no longer dimensionless and have dependence on the
continuous recoil energy $T_r$, the creation of an
ionized state can still be defined as
$\ket{T_r \kappa m_j} 
\equiv
    a^\dagger_{T_r \kappa m_j} \ket{\Omega_0}$.
One may borrow some experience from the free-particle
quantization \cite{Peskin:1995ev} where a momentum
eigenstate is defined as
$|\bm p\rangle \equiv \sqrt{2 E_{\bm p}} a^\dagger_{\bm p} |0\rangle$
with the prefactor $\sqrt{2 E_{\bm p}}$
added to keep
Lorentz invariance. Although the Lorentz
invariance is no longer true for an atomic system
with the nucleus assigned to be at the coordinate origin,
the three-dimensional rotation invariance is still kept.
For an ionized electron, its asymptotic momentum $\bm p$
should have a measure $d^3 \bm p$ in the three-dimensional
phase space and the corresponding $\delta$ function is
$\delta^{(3)}(\bm p)$. However, only the recoil energy
$T_r$ or equivalently the asymptotic momentum magnitude
$|\bm p|$ can be experimentally reconstructed.
It is then necessary to reduce the phase-space element
and the corresponding $\delta$ function to $d |\bm p|$
and $\delta(|\bm p|)$. In the polar coordinate system,
the three-dimensional $\delta$ function should take the
form as,
$\delta^{(3)}(\bm p) = \frac 1 {|\bm p|^2} \delta(|\bm p|) \delta(\varphi) \delta(\cos \theta)$
with integration over the momentum size $|\bm p|$,
the zenith $\theta$ and azimuthal $\varphi$ angles,
to maintain the defining property of a
$\delta$ function, $\delta^{(3)}(\bm p) |\bm p|^2 d |\bm p| d \varphi d \cos \theta = 1$.
In other words, the $\delta$ function and 
phase space integration can reduce to
$\delta(|\bm p|) d |\bm p|$. Considering the
relation between $T_r$ and $\bm p$,
$\partial |\bm p| / \partial T_r = (T_r + m_e) / |\bm p|$.
So the reduction further advances to $\delta(T_r) d T_r$.
Although this final form can be guessed from naive
expectation, our derivation makes explicit that it is
fully consistent with the exact three-dimensional phase space.
Correspondingly, the orthogonality of ionized states
becomes, $\langle T'_r \kappa' m'_j | T_r \kappa m_j \rangle =
2 \pi \delta_{\kappa \kappa'} \delta_{m_j m'_j} \delta(T_r - T'_r)$.

Similar to the bound states, the ionized electron field
is also constructed from the corresponding annihilation
operator $a_{T_r \kappa m}$ and wave function
$\psi_{T_r \kappa m_j} ({\bm x})$,
\begin{equation}
  e_I(x)
\equiv 
\sum_{\kappa m_j} 
\int \frac{d T_r }{2 \pi} a_{T_r \kappa m_j} 
\psi_{T_r \kappa m_j}(\bm{x}) 
e^{-i T_r t}.
\label{ionized_field}
\end{equation}
Since the final-state kinetic energy forms a continuous
spectrum, the summation over states is replaced by a
one-dimensional integral, $\int dT_r/2 \pi$.
The wave function can also be obtained by applying
the field operator on the particle state as
$e_I(x) \ket{T_r \kappa m_j} =
e^{-i T_r t}\psi_{T_r \kappa m_j} (\bm x) \ket{\Omega_0}$.

According to quantum mechanics, the normalization
condition for continuum wave functions is
\cite{2007rqta.book.....G},
\begin{equation}
  \int 
  \psi^\dagger_{T_r \kappa m_j}
  ({\bm x}) 
  \psi_{T'_r \kappa 'm'_j}({\bm x}) 
  d^3 {\bm x}
=
  2 \pi
  \delta_{\kappa \kappa'} 
  \delta_{m_j m'_j} 
  \delta( T_r - T'_r ).
\label{normFinal}
\end{equation}
So the wave function has a dimension of
$[\psi_{T_r \kappa m_j} (\bm x)] = 1$.
Since the mass dimension of the annihilation
operator is $[a_{T_r \kappa m_j}] = -1/2$,
the field operator has a dimension of
$[e_I (\bm x)] = 3/2$ that is the same
as the bound state and free particle cases.

\subsection{Phase Space and Cross Section}
\label{sec:xsec}

From the above derivation of the ionized electron
field, one may see that the final-state phase space
integration is,
\begin{align}
  d \Omega'_e
=
  \sum_{\kappa m_j}
  \int \frac {d T_r}{2 \pi},
\end{align}
which should enter \geqn{eq:PS}.
The dimensionality of this phase space
integration $[d \Omega'_e] = 1$ and its
relation to the dimensionality of the
$\mathcal{T}$-matrix
$[\mathcal{T}] = - [d \Omega'_e]/2 - 2 = - 5/2$
as established
below \geqn{decoupleT} are self-consistent.
Since the dimensionality of the $\mathcal T \sim \braket{{\bm p}'_{\chi},f|{\bm p}_{\chi},i}$
is given by the sum of the dimensionalities
of four scattering quantum states, it is also $[\mathcal{T}] = -5/2$
as the free DM state has a dimension of $[\ket{{\bm p}_\chi ({\bm p}'_\chi)}] = -1$ and the initial (final) state of electron has a dimension of $[\ket{i (f)}] = 0 (1/2)$.
It can also be observed that the wave function normalization
with a $2 \pi$ factor in \geqn{normFinal}
does not affect the calculation of scattering cross section
since it cancels out with the $1/2 \pi$ factor
in the phase space element.

Then, the
scattering cross section in \geqn{CSnkm} becomes
\begin{equation}
\sigma^{ion}_{n \kappa} 
= 
    \frac{1}{2 |{\bm v}_\chi| E_{{\bm p}_{\chi}}}
    \sum_{\kappa' m'_j m_j} \int \frac{d T_r}{2 \pi}
  \int \dFs{{\bm p}_{\chi}'} 
  \overline{|\mathcal{M}({\bm p}_{\chi},{\bm p}_{\chi}')|^2}
  (2\pi) \delta(E_{{\bm p}_{\chi}'} - E_{{\bm p}_{\chi}} + \Delta E),
\label{CSnkm_w_ps}
\end{equation}
as the total cross section for the
DM scattering with all those initial electrons
with quantum numbers $n$ and $\kappa$.
Note that the spin average is defined as
$\overline{|\mathcal{M}({\bm p}_{\chi},{\bm p}_{\chi}')|^2}
\equiv \sum_{s_\chi} |\mathcal{M}({\bm p}_{\chi},{\bm p}_{\chi}')|^2/J_\chi$
for the scattering with a single specific eletron,
with the average performed over only the DM spin.
The spin degree of freedom $J_\chi = 2$ for a fermionic DM.
In the case of free-electron scattering with the
spatial wave function being a plane wave, this spin
average should also been performed for the initial electron.
However, for a bound electron inside atom, the initial
wave function is in general a function of the electron
spin which renders the spin average non-trivial and has
been incorporated in $m_j$
\cite{Dyall2007IntroductionTR,2007rqta.book.....G}. The
only exception is the non-relativistic limit with
possible factorization or detachment of the spatial wave function
and the spinor. These details will be elaborated
in the following \gsec{sec:non_rel_f} and
\gsec{sec:relativisticEffects}.

The detector in DM direct detection experiments 
is capable of measuring only the energy transfer between the initial and final electrons, which is nearly independent of magnetic quantum number $m_j$. 
That is why we sum over $m_j$ in 
the cross section level. 
The differential cross section with respect to recoil energy $T_r$ is then, 
\begin{equation}
 \frac{d \sigma^{ion}_{n \kappa }|{\bm v}_\chi|}
 {d T_r}
= 
    \frac{1}{4 \pi E_{{\bm p}_{\chi}}}
    \sum_{\kappa' m'_j m_j} 
  \int \dFs{{\bm p}_{\chi}'}
  \overline{|\mathcal{M}({\bm p}_{\chi},{\bm p}_{\chi}')|^2}
  (2\pi) \delta(E_{{\bm p}_{\chi}'} - E_{{\bm p}_{\chi}} + \Delta E),
\label{Xsec1}
\end{equation}
for the total contributions from all
those initial electrons with quantum number $\kappa$.

The integration over $d^3 {\bm p}'_\chi$ can be further transformed into 
$d^3 {\bm q}$ 
using the relation ${\bm q} \equiv {\bm p}_\chi - {\bm p}'_\chi$. 
Then, the integration over the scattering angles, $\theta_q$ and $\varphi_q$, 
can be performed by 
leveraging the 
kinematics provided by the energy 
conservation \geqn{eq:Echi}, 
\begin{eqnarray}
\int
  \delta
\left(
  \Delta E -E_{{\bm p}'_\chi} + E_{{\bm p}_\chi}
\right)
  d^3 {\bm q}
= 
    \int
  \frac{2 \pi E_{{\bm p}'_\chi} |{\bm q}| d|{\bm q}|}
       {|{\bm p}_\chi| }.
\label{eq:delta-cosTheta}
\end{eqnarray}
The dependence on the
azimuthal angle $\varphi_q$ is isotropic 
and its integration yields a factor of $2 \pi$.
Finally, the differential cross section still
contains the integration over the momentum
transfer magnitude $|{\bm q}|$ as, 
\begin{equation}
  \frac {d \sigma^{ion}_{n \kappa }|{\bm v}_\chi|}
        {d T_r}
= 
  \sum_{\kappa' m'_j m_j} 
  \frac{1}{16 \pi^2 |{\bm p}_\chi | E_{{\bm p}_\chi}}
  \int_{{|\bm q|}_{\min}}^{{{|\bm q|}_{\max}}}
  \overline{|\mathcal{M}({\bm q})|^2}
  |{\bm q}| d|{\bm q}|,
\label{eq:dSigmadT}
\end{equation}
with the integration limits shown in \geqn{eq:q-limits}. 
Due to a broad momentum distribution of those electrons
in the atomic Coulomb potential, the cross section of
being ionized to a fixed recoil energy should include all
the contributions from various momentum transfers.

\section{Factorization of Scattering Matrix in Non-Relativistic Limit}
\label{sec:factorization}

In the above derivation of the scattering cross section,
we have only formally introduced the $\mathcal T$ and
$\mathcal M$ matrices in \geqn{Tmatrix} and \geqn{decoupleT},
respectively. However, in practical calculations, it is
necessary to specify their concrete forms.
In this section, we will derive the explicit expressions
for $\mathcal T$ and $\mathcal M$. 
The electron contribution to the scattering matrix element
$\mathcal M$ can be factorized into a product of a
spatial wave function part and a spinor part in the
non-relativistic limit. Most importantly, it can be made
explicit that the negative-value problem elaborated
in \gsec{sec:Factorization} can be naturally resolved.

\subsection{Matrix Element}

With a well-defined electron field combining both the bound
and ionized states in \gsec{sec:bound} and \gsec{sec:ionized},  
\begin{equation}
   e (x)
=
    \sum_{n \kappa m_j} 
    a_{n \kappa m_j}  \psi_{n \kappa m_j}(\bm{x}) e^{-i E_{n \kappa} t}
 +
    \sum_{\kappa m_j} 
    \int 
    \frac{dT_r}{2\pi} 
    a_{T_r \kappa m_j} 
    \psi_{T_r \kappa m_j}(\bm{x}) 
    e^{-i T_r t},
\label{ElectronField}
\end{equation}
one can write down the Lagrangian for the interaction
between DM and electron to perform explicit calculations.
To further elaborate the formalism while maintaining
generality, we assume the DM-electron interaction is
mediated by an intermediate particle $\phi$, 
\begin{equation}
    \mathcal{L}_{\rm eff} 
= 
    \bar \chi \Gamma_A \chi \phi^A  
+ \bar e \Gamma_B e \phi^B.
\end{equation}
The indices $A$ and $B$ stand for different types of
Lorentz structure for generality. Note that $A$ and
$B$ are dummy indices that need not to be summed.
In this work, we assume a heavy-mediator whose mass is
much larger than the typical scattering momentum transfer,
such that the interaction between electron and DM can
be written as a contact operator
$\mathcal{L}_{\text{eff}} \sim \bar \chi \Gamma_A \chi \bar e \Gamma^A e / \Lambda^2$
with $\Lambda$ being the cutoff scale.

The $\mathcal{T}$ matrix \geqn{Tmatrix} 
in the position space is then simply, 
\begin{equation}
  \mathcal T
\equiv
    \bra{f , {\bm p}'_{\chi}} 
    \int d^4 x \bar{\chi} (x) \Gamma_A \chi (x) \phi^A(x) 
    \int d^4 y \bar{e}(y) \Gamma_B e (y) \phi^B (y) 
    \ket{i,{\bm p}_{\chi}}.
\end{equation}
As elaborated in \gsec{sec:bound} and \gsec{sec:ionized},
$\ket i$ is the initial bound state $\ket{n \kappa m_j}$ and
$\ket f$ is the final ionized one $\ket{T_r \kappa' m'_j}$.
Applying the DM field operator $\chi$ on the initial and
final states as well as contracting the mediator fields
into a propagator, the $\mathcal T$ matrix element becomes 
\begin{equation}
    \mathcal{T}
=
  \int d^4 x 
  \bar u_{\chi}({\bm p}'_{\chi}) e^{i p'_\chi \cdot x}
  \Gamma_A
  u_\chi ({\bm p}_{\chi}) e^{- i p_\chi \cdot x}
\int \frac {d^4 q}{(2 \pi)^4}
D_{AB}(q) e^{i q \cdot (x - y)}
  \int d^4 y \bra{f}
  \bar{e}(y) \Gamma_B e(y) \ket{i},
\end{equation}
where $D_{AB}$ generally denotes the $\phi$ propagator.

The integration over the DM position ($x$) imposes an
energy-momentum conservation condition on the DM momentum
transfer $q \equiv p_\chi - p'_\chi$. In addition,
the electron field operator extracts out the electron wave
function $e^{-i E_i t}\psi_i({\bm y})$
($e^{-i E_f t}\psi_f ({\bm y})$) from the initial (final)
states to give
\begin{equation}
    \mathcal T
=
    \bar u_\chi({\bm p}_{\chi}') 
    \Gamma_A u_\chi ({\bm p}_{\chi})
    D_{AB} (q)
    \int d^4 y 
    e^{- i q \cdot y}
    e^{i (E_f - E_i) t} 
    \bar \psi_{f} ({\bm y}) 
    \Gamma_B
    \psi_{i} ({\bm y}).
\end{equation}
Note that the electron wave functions $\psi_i(\bm y)$
and $\bar \psi_f(\bm y)$ only depend on the spatial
coordinate $\bm y$. It is then possible to first
perform the time integration to give the energy
conservation $\delta$ function 
\begin{equation}
\mathcal T
=
    \bar u_\chi({\bm p}_{\chi}') \Gamma_A u_\chi ({\bm p}_{\chi})
    D_{AB} (q)
    \int d^3 {\bm y}
    e^{i {\bm q \cdot y} }
    \bar \psi_{f} ({\bm y}) 
    \Gamma_B
    \psi_{i} ({\bm y}) 
    (2 \pi)
  \delta \left( \Delta E - E_{{\bm p}_\chi} + E_{{\bm p}'_\chi} \right).
\end{equation}
The corresponding $\mathcal M$ matrix element is the
part after extracting the energy conservation $\delta$
function as already defined in \geqn{decoupleT}, 
\begin{equation}
\mathcal{M}({\bm p}_{\chi},{\bm p}_{\chi}') \equiv 
\bar u_\chi({\bm p}_{\chi}') \Gamma_A u_\chi ({\bm p}_{\chi})
    D_{AB} (q)
    \int d^3 {\bm y}
    e^{i {\bm q \cdot y} }
    \bar \psi_{f} ({\bm y}) 
    \Gamma_B
    \psi_{i} ({\bm y}).
\label{mdefine}
\end{equation}
We have now reached the key distinction between the
scatterings with a free or bound electron target. 
In the presence of a Coulomb potential, the electron
wave functions have a more complicated dependence on
the spatial coordinate $\bm y$, rather than an exponential 
factor as in the free particle case. Therefore, 
the integration over ${\bm y}$ can not be 
easily performed and only the energy-conserving
$\delta$ function can be extracted.

\subsection{Atomic Form Factor in Non-Relativistic Approximation}
\label{sec:non_rel_f}

In the scattering matrix element of \geqn{mdefine},
the DM and electron parts have been factorized.
It is possible to further factorize the spatial
wave and spinor wave functions for the electron bilinear
as we will elaborate in this subsection.

In the relativistic quantum mechanics, the electron
wave function is determined by the covariant Dirac
equation,
\begin{equation}
\left[
  i \slashed \partial
- V(|{\bm r}|) \gamma^0
- m_e
\right] \psi_\alpha (\bm r, t) 
=
  0,
\label{FieldEquation}
\end{equation}
in the presence of the atomic Coulomb potential $V(|{\bm r}|)$.
With the hydrogen-like atom approximation \cite{Bethe:1957ncq},
the Coulomb potential 
is $ V(|{\bm r}|) \equiv - Z_{n\kappa} e^2 /4 \pi |{\bm r}|$ with 
an effective charge $Z_{n\kappa}$ that varies with the energy level
labeled with $n$ and $\kappa$. For convenience, we have
switched the electron coordinate $\bm y$ in \geqn{mdefine} to
$\bm r$ which is more commonly used for central potential,
in the later part of this paper.

For a stationary state, its wave function can be decoupled
into spatial and temporal components as 
$\psi_\alpha (\bm r, t) \equiv \psi_\alpha ({\bm r}) e^{-i E_\alpha t}$.
In such a way, the time-independent Dirac equation becomes, 
\begin{equation}
  \gamma^0
  (i{\bf \boldsymbol \gamma \cdot \boldsymbol \nabla} + m_e)
  \psi_\alpha ({\bm r})
=
  [E_\alpha - V(|\bm r|)] \psi_\alpha ({\bm r}),
\label{eq:dirac}
\end{equation}
with these $\gamma$ matrices defined in the chiral representation.
The wave function $\psi_\alpha$ is a four-component spinor.
One can perform a small-momentum expansion in the
non-relativistic limit to investigate the form of
the scattering matrix element \cite{Weinberg:1995mt},
\begin{equation}
		\psi_{a,m_s} (\bm{x})
		\approx
		\frac{1}{\sqrt{2}}
		\left\lgroup
		\begin{matrix}
			(1 + i {\boldsymbol \sigma \cdot \boldsymbol \nabla} / 2 m_e) f_{a,m_s} (\bm{r}) \\
			(1 - i {\boldsymbol \sigma \cdot \boldsymbol \nabla} / 2 m_e) f_{a,m_s} (\bm{r})
		\end{matrix}
		\right\rgroup,
		\qquad
		f_{a,m_s} (\bm{r})
		\equiv
		\phi_a (\bm{r}) \eta_{m_s},
		\label{DiracSolution1}
\end{equation} 
where the two-component spinor $\eta^T_{m_s} = (1,0)$ is
for the spin-up state with spin magnetic quantum number
$m_s = +1$ and $(0,1)$ for
the spin-down state with $m_s = -1$. 
The original quantum number $a \equiv (n \kappa m_j)$ for the relativistic case
is now decomposed into the spin $m_s$ and the quantum number $a \equiv (nl m_l)$,
which corresponds to the spatial part with $l$ being the
angular momentum and $m_l$ being its magnetic quantum number. Since the spinor
$\eta_{m_s}$ is dimensionless, the spatial wave function
is of dimension $3/2$ for the initial bound state
or $1$ for the final ionized state as determined by
their normalization conditions in \geqn{eq:bound-norm}
and \geqn{normFinal}, respectively.

In the non-relativistic limit, namely omitting the
$\boldsymbol \sigma \cdot \boldsymbol \nabla$ term
in \geqn{DiracSolution1}, the spinor $\eta_s$ can
entirely decouple from the spatial wave function
$\phi_a(\bm r)$. 
Then the spatial component $\phi_a$ then satisfies
the Schrodinger equation
\cite{Essig:2012yx,Catena:2019gfa,Ge:2021snv},
\begin{equation}
- \frac{\nabla^2}{2 m_e} \phi_a({\bm r})
+ V(|{\bm r}|) \phi_a({\bm r})
= E_a \phi_a ({\bm r}).
\label{eq:Schrodinger}
\end{equation}
In the approximation with a hydrogen-like atom,
the bound-state solution is characterized by the
principal quantum numbers $n$, orbital angular momentum $l$,
and magnetic quantum numbers $m_l$ while for the
ionized-state solution, the principal quantum number
$n$ is replaced by the electron recoil energy $T_r$. 

Neglecting the momentum corrections encoded
in the spatial-derivative terms, the wave function
\geqn{DiracSolution1} can be decomposed into a
spatial wave function $\phi_a$ and the electron
spinor $u(m_e)$ as $\Psi \equiv u(m_e) \phi/\sqrt{2 m_e}$
at rest \cite{Ge:2021snv}. As a result, the inner
product between the initial and final states
appearing in the matrix element can factorize out as,
\begin{equation}
  \int d^3 {\bm r}
  e^{i {\bm q} \cdot {\bm r} }
  \bar \psi_{f} ({\bm r}) \Gamma_B \psi_{i} ({\bm r})
=
  \frac 1 {2 m_e} \bar u_{s'}(m_e) \Gamma_B u_s(m_e)
  \int d^3 {\bm r}
  e^{i {\bm q} \cdot {\bm r} }
  \phi^*_{f} ({\bm r}) \phi_{i} ({\bm r}).
\label{innerproduct}
\end{equation}
Since the 
coupling matrix $\Gamma_B$ in the spinor space
can only contract with the electron spinor $u(m_e)$, it can be
extracted from the spatial integral.
Note that the spinor part has become a constant in the
non-relativistic limit. It is much more convenient to express
the spinor part in the usual four-component spinor form with
a prefactor $1/2m_e$ to balance the mass dimension.
Consequently, the scattering $\mathcal M$ matrix factorizes
as \cite{Ge:2021snv}, 
\begin{equation}
  \mathcal{M}
\equiv
  \frac{1}{2 m_e} \mathcal{M}_0 ({\bm q}) 
\times 
  \int d^3 {\bm r} 
  e^{i {\bm q} \cdot {\bm r}}
  \phi^*_{E_r l' m'_l} ({\bm r}) 
  \phi_{n l m_l} ({\bm r}).
\label{MM0}
\end{equation}
Being constructed with $u(m_e)$ which contains
only the electron mass $m_e$ but no momentum,
the matrix element,
\begin{equation}
		\mathcal{M}_0
		\equiv 
		\bar u_\chi({\bm p}_{\chi}') \Gamma_A u_\chi ({\bm p}_{\chi})
		D_{AB} (q)
		\bar u(m_e) \Gamma_B u (m_e),
\label{M_nr}
\end{equation}
is for the scattering with an electron at rest. 
Depending on the Lorentz structure, the electron
bilinear part takes the form as \cite{Ge:2021snv},
\begin{eqnarray}
  \frac 1 {2 m_e} \bar u(m_e) \Gamma_B u(m_e)
\equiv
  \bar u(m_e)
\left\{
  \frac 1 {2 m_e},
  \frac {\gamma_0 {\bf q} \cdot \boldsymbol \gamma \gamma^5} {4 m^2_e},
  \frac {\gamma^0} {2 m_e},
  \frac {\gamma^5 \gamma^i} {2 m_e},
  \frac {[\gamma^i, \gamma^j]} {2 m_e}
\right\}
  u(m_e).
\label{nr_freem}
\end{eqnarray}
The remaining part is the so-called atomic form factor
\cite{Essig:2011nj,Essig:2012yx,Essig:2015cda},
\begin{equation}
  f^{T_r l'm'_l}_{nlm_l}({\bm q})
\equiv
  \int d^3 \bm r
  e^{i\bm q \cdot \bm r}
  \phi_{T_r l'm'_l}^{*} (\bm r) 
  \phi_{n l m_l} (\bm r).
\label{formfactor-1}
\end{equation}

With the scattering matrix defined in this way,
we can write down the differential scattering
cross section according to \geqn{eq:dSigmadT},
\begin{equation}
  \frac {d \sigma^{ion}_{n l}|{\bm v}_\chi|}
        {d T_r}
= 
  \frac 1 {16 \pi^2 |{\bm p}_\chi | E_{{\bm p}_\chi}}
  \int_{{|\bm q|}_{\min}}^{{{|\bm q|}_{\max}}}
  \frac{\sum_{s_\chi,s'_\chi} \sum_{s_e,s'_e} |\mathcal{M}_0|^2}{J_\chi (2m_e)^2}
  \sum_{l' m_{l'} m_l} 
  \left|f^{T_r l'm'_l}_{nlm_l}({\bm q})\right|^2
  |{\bm q}| d|{\bm q}|,
\label{eq:dSigmadT_nr}
\end{equation}
as the total contribution from all those
electrons with quantum numbers $n$ and $l$.
Since the non-relativistic quantum numbers $(n,l,m_l)$
do not contain the electron spin information \cite{cohen1977quantum},
the scattering matrix part $|\mathcal M|^2$
has been summed and averaged
over the initial spin states of both DM ($s_\chi$)
and electron ($m_s$). So the spin-averaged
scattering matrix element part can be defined as
$\overline{|\mathcal M_0|^2} \equiv \sum_{s_\chi,s'_\chi} \sum_{m_s,m'_s} |\mathcal{M}_0|^2 / 2 J_\chi$
for the scattering with a single electron where $m_s (m'_s)$ and $s_\chi (s'_\chi)$ are the spin magnetic quantum number for the initial (final) electron and DM states.
However, there are two electrons with the opposite spin
for given $(n,l,m_l)$ and consequently the
factor $2$ that should appear in the denominator
is compensated with only $J_\chi$ appearing in
\geqn{eq:dSigmadT_nr}.

Furthermore, the summation over the final-state angular momentum
($l'$) and the magnetic quantum numbers ($m_{l'}, m_l$)
only acts on the form factor, which can be altogether
defined as the non-relativistic $K$-factor of
initial-state $\ket{nl}$,
\begin{equation}
    K_{nl} \equiv 
    \frac{1}{2l+1}
    \sum_{l' m_{l'} m_l} 
    \left|
    \int d^3 \bm r
  e^{i\bm q \cdot \bm r}
  \phi_{T_r l'm'_l}^{*} (\bm r) 
  \phi_{nlm_l} (\bm r)
    \right|^2.
\label{K_nr}
\end{equation}
Due to the summation over $m_l$, there are $(2l+1)$ electrons in the quantum state with angular momentum $l$ \cite{cohen1977quantum}. Averaging over these electrons yields the atomic factor $K_{nl}$ for a single electron in the state $\ket{nl}$.

With \geqn{MM0}, one can recalculate the matrix element
of DM-electron scattering with vector interactions.
Since the electron spinor $u(m_e)$ at rest no longer carriers momentum
dependence, the free-electron scattering matrix element
\geqn{Mdp} in the literature \cite{Cao:2020bwd} should
be modified as,
$\sum_{s_\chi} \sum_{m_s}|\mathcal{M}_0|^2
=
    64 m_\chi 
    m_e (m_\chi + T_\chi)^2
    /(t -m_{A'}^2)^2$ with $m_{A'}$ being the dark photon mass,
which is positive definite to naturally resolve the negative
value issue that arises in the calculation with a free
electron as detailed in \gsec{sec:Factorization}.
In other words, the scattering matrix element
between DM and bound electrons cannot be simply factorized
into the product of a free-electron scattering matrix element
and an atomic form factor. The factorization can be achieved
only under the non-relativistic approximation through the
expansion of the wave function. Effectively, this puts the
electron at rest rather than allowing it to move freely. 
This approach has also been applied in recent work \cite{Alhazmi:2025nvt}.
% However, the underlying assumption is still not fully self-consistent within a relativistic formalism.

\section{Relativistic Effects}
\label{sec:relativisticEffects}

The key difference between the scattering with a
free electron and an electron in a Coulomb potential
lies in their matrix element \geqn{mdefine}.
Although we have demonstrated the factorization
of the spatial and spinor wave functions of an
atomic electron in the non-relativistic limit
in \gsec{sec:non_rel_f},
a more realistic calculation should not
limit the kinematic space. In this section,
we will explore the relativistic case and
make comparison with its nonrelativistic
counterpart. Moreover, the formalism and relativistic effects discussed in this section apply across the full kinematic parameter space of DM. We will no longer restrict ourselves to CRDM or the non-relativistic approximation. 

The spin-averaged matrix element square reads,
\begin{equation}
\begin{split}
    \overline{|\mathcal{M}({\bm q})|^2}
& =
  \frac{1}{J_\chi}
\tr[ (\slashed{p}_\chi + m_\chi) \Gamma_A
    (\slashed{p'}_\chi + m_{\chi'}) \Gamma'_A ]
    |D_{AB} ({\bm q})|^2 
\\
& \times
\tr
\left[ \int d^3 {\bm r}
    e^{ i {\bm q \cdot r} }
    \bar \psi_{f} ({\bm r}) 
    \Gamma_B
    \psi_{i} ({\bm r}) 
    \int d^3 {\bm r}
    e^{- i {\bm q \cdot r} }
    \bar \psi_i ({\bm r}) 
    \Gamma'_B
    \psi_f ({\bm r})
\right],
\end{split}
\label{eq:Msq}
\end{equation}
with summation and average over the DM spins.
Since the electron field
does not contain a free fermion spinor $u({\bm p}_e)$,  
one cannot easily perform spin summation on them to obtain the conventional 
$\slashed{p}_e + m_e$ structure. 
Especially, the spin 
$s_e$ is no longer a good quantum number which makes
the summation or average over it meaningless.
Instead, the information of spin has been
incorporated into the magnetic quantum number $m_j$.
Consequently, the subsequent trace technique is naturally inapplicable to the bound-electron case and the 
squared matrix element can only be expressed by the inner product of the initial and final wave functions.
The initial electron state is fixed and the
corresponding wave function is then used for evaluation.
So one can perform spin average only for the initial DM
particle, resulting in a factor of $1/J_{\chi}$. 
The spin-averaged matrix element 
$\overline{|\mathcal{M}({\bm q})|^2}$ has
three components in \geqn{eq:Msq}, the DM part
$|\mathcal M^{AA}_\chi|^2
\equiv 
\tr[ (\slashed{p}_\chi + m_\chi) \Gamma_A
    (\slashed{p'}_\chi + m_{\chi'}) \Gamma'_A ]
/ J_\chi$, the propagator $|D_{AB} ({\bm q})|^2$, and 
the electron part
with the inner product of the initial and final wave functions with the Lorentz factor $\Gamma_B$ ($\Gamma'_B$).

To compare the electron part with the scalar-type atomic factor in literature
\cite{Roberts:2016xfw,Ge:2021snv}, we take
$\Gamma_S = 1$
for illustration. Then the four-component
spinor function $\psi_{i(f)}$ contracts with 
the gamma matrix $\gamma^0$ into a scalar 
function and the three defined parts in \geqn{eq:Msq}
above can 
be completely decoupled. Consequently, the
differential cross section \geqn{eq:dSigmadT}
becomes, 
\begin{align}
  \frac{d \sigma^{S}_{n \kappa }|{\bm v}_\chi|} {d T_r}
= 
  \frac{2|\kappa|} {16 \pi^2}
  \int_{{\bm q}_{\min}}^{{{\bm q}_{\max}}}
  \frac {|{\bm q}| d|{\bm q}|}
        {|{\bm p}_\chi | E_{{\bm p}_\chi}}
\overline{|M^{SS}_\chi|^2}
  D_{SS}^2 (q) 
  K_{n \kappa}^{S} ({\bm q}, \Delta E),
\label{dcsWithK}
\end{align}
where $K^S(n \kappa)$ is the so-called $K$-factor.
We have combined the summation over the initial-
and final-state quantum numbers $(\kappa', m'_j, m_j)$
with the matrix element of the electron part
into the $K$-factor,
\begin{equation}
  K_{n \kappa}^{S} ({\bm q}, \Delta E)
\equiv
  \frac{1}{2|\kappa|}
    \sum_{\kappa' m'_j m_j}
\left|
  \int d^3 {\bm r}
    e^{i {\bm q \cdot r} }
    \psi^\dagger_{T_r \kappa' m'_j} ({\bm r}) 
    \gamma^0
    \psi_{n \kappa m_j} ({\bm r})
\right|^2.
\label{eq:K0}
\end{equation}
For the convenience of comparison, similar to
the non-relativistic case, one can average over
the summation of the initial-state magnetic
quantum number $m_j$ with the $1/2|\kappa|$ prefactor
to obtain the $K$-factor for a single electron.
A given state 
$\ket{n \kappa m_j}$ with quantum number $\kappa$
fixes the total angular momentum $j$ and hence
the additional factor $2 |\kappa| = 2j+1$
is actually the degree of freedom for the initial state.
The relationship
between the quantum numbers $j$ and $\kappa$ will be discussed later in \gsec{sec:relSpinor}.
Since the average of the initial states is
taken in the $K$-factor, the same $2 |\kappa|$ factor is
multiplied in the differential cross section
\geqn{dcsWithK} for compensation.

Using the Wigner–Eckart theorem, \geqn{eq:K0} can be further simplified to,
\begin{equation}
  K^S_{n,\kappa} (\Delta E, {\bm q})
=
    \frac{1}{2|\kappa|}
  \sum_{\kappa'} \sum_{L}
    \frac{1}{4 \pi} 
    \begin{pmatrix}
      j' & L & j \\
      -\frac12 & 0 & \frac12 
    \end{pmatrix}^{-2}
    \left|
\left\langle
  T_r \kappa' \left. \frac12 
  \right| \gamma^0_D T_{L 0}
  \left| n \kappa \frac 1 2 \right.
\right\rangle
    \right|^2,
\label{kernel3}
\end{equation}
where the parentheses denote the Wigner-$3j$ symbol The subscript $D$ denotes the Dirac representation of the $\gamma$ matrices since the atomic wave functions are also solved in the Dirac representation as will be shown later.
After simplification, the original triple summation reduces to a double summation, and the inner product of initial and final wave functions now incorporates only states with magnetic quantum number $m_j = 1/2$. 
The details of simplification are summarized in \gapp{Wigner_theorem}.

This electron kernel function, or $K$-factor \cite{Roberts:2016xfw,Ge:2021snv},
(sometimes electron form factor, or $f$-factor \cite{Essig:2011nj}) describes the 
atomic effects in DM-bound electron scattering process.
By taking the scalar interaction as an example, we will simplify and calculate this theoretically
consistent $K$-factor
and study its relativistic effects with the Dirac
spinor wave functions in this section.

\subsection{Spinor Solutions of Dirac Equation and the Inner Product of Radial Wave Functions}
\label{sec:relSpinor}

To proceed with the relativistic calculation of the $K$-factor, we must explicitly define the specific forms of the wave functions involved in the inner product of \geqn{kernel3}.
The relativistic spinor wave function of electrons within the Comloub potential are the solutions of Dirac equation \geqn{eq:dirac}, but in the Dirac representation.
As mentioned in \gsec{subsec:2ndQuantization},
the angular parts of this isotropic Dirac equation are the same 
for both the bound and ionized states. 
Consequently, the four-spinor solutions of the Dirac equation 
\geqn{FieldEquation} for states $\ket{n (T_r) \kappa m_j}$ \cite{Dyall2007IntroductionTR,2007rqta.book.....G}, 
\begin{equation}
  \psi_{n (T_r) \kappa m_j}(|\bm r|, \Omega)
=
  \frac 1 {|\bm r|}
\left(
\begin{array}{c}
    a P_{n (T_r) \kappa}(|\bm r|) 
    \sqrt{\frac{l +\frac{1}{2} + a m_j}
     {2 l +1 }}
     Y_{l}^{m_j-1 / 2} 
\\
     P_{n (T_r)  \kappa}(|\bm r|) 
     \sqrt{\frac{l +\frac{1}{2} - a m_j}
     {2 l + 1}} Y_{l}^{m_j+1 / 2}
\\
    - a i Q_{n(T_r)  \kappa}(|\bm r|) 
    \sqrt{\frac{l_s +\frac{1}{2} - a m_j}
     {2 l_s + 1}}
     Y_{l_s}^{m_j-1 / 2}
\\
     i Q_{n (T_r) \kappa}(|\bm r|) 
     \sqrt{\frac{l_s +\frac{1}{2} + a m_j}
     {2 l_s + 1}} Y_{l_s}^{m_j+1 / 2}
\end{array}
\right),
\label{wavefunction_general}
\end{equation}
with $a \equiv -\mbox{Sign}(\kappa)$ being the opposite sign
of $\kappa$, share the same angular wave functions. 
The only difference appears in the radial part
with $P_{n\kappa}$ and $Q_{n \kappa}$ for the
bound electron case while $P_{T_r \kappa}$ and
$Q_{T_r \kappa}$ for the ionized one,
with the principle quantum number $n$ replaced
by the electron recoil energy $T_r$.

The total angular momentum $j$ of one electron, 
which is a good quantum number, 
comes from the coupling between the orbital
angular momentum $l$ and spin $s = 1/2$. 
The absolute value of $\kappa$ corresponds to the
total angular momentum $j$ by $j = |\kappa| - 1/2$,
while the sign of $\kappa$ indicates how the
orbital angular momentum $l$ couples with spin $s$
which is always $1/2$. 
The quantum number $\kappa$ is positive 
for $l = j + 1/2$ and negative for $l = j - 1/2$.
Equivalently, $j = l + \frac a 2$ applies for both cases.
The single variable $\kappa$ determines both
the orbital angular momentum $l$ and the total
one $j$.
Given state $\ket{n(T_r) \kappa m_j}$, the angular wave function of the upper two-spinor has quantum numbers $\kappa$ and $m_j$, while the lower two-spinor has the quantum numbers $-\kappa $ and $m_j$
due to the parity structure of the Dirac four-spinor
\cite{Dyall2007IntroductionTR}. 
As a result, the orbital angular momentum $l_s$
of the lower two-component spinor is different 
from the upper one $l$ by 1,
$l = l_s + a$. The subscript $s$ means that
this angular momentum $l_s$ is associated
with the small-component radial wave function $Q$
while $l$ is for the large component $P$.

\subsubsection{Bound Radial Wave Function}

The radial wave function of a bound electron
state $\ket{n \kappa m}$
\cite{Dyall2007IntroductionTR} is, 
\begin{align}
    P (Q)_{n \kappa} (|\bm r|)
& =
    \frac{N}{2 Z_{n \kappa}}
    \frac{\sqrt{ \Gamma (2 \gamma +n' + 1)}}{\Gamma(2\gamma + 1) \sqrt{n' !}}
    \sqrt{\frac{1 \pm \epsilon}{4 N (N - \kappa)}}
    \left( \frac{2 Z_{n \kappa}}{N}\right)^{\frac32} 
    e^{- \frac{Z_{n \kappa} |\bm r|}{N}}
    \left(\frac{2 Z_{n \kappa} |\bm r|}{N} \right)^{\gamma} 
\label{Rel_Bound_Wave_L}
\\
& \times 
\left[ 
  \mp n' {}_1F_1 \left(-n'+1, 2\gamma + 1, \frac{2 Z_{n \kappa} |\bm r|}{N}\right)
  +
  (N - \kappa) {}_1F_1 \left(-n' , 2\gamma +1, \frac{2 Z_{n \kappa} |\bm r|}{N}\right)
\right],
\nonumber
\end{align}
with the upper sign for the large component $P$
and the lower sign for the small one $Q$.
In this expression, ${}_1F_1$ is the confluent
hypergeometric function and $Z_{n \kappa}$ is the effective charge.
And the other parameters are defined by the principle
($n$) and angular ($l$) quantum numbers, 
\begin{equation}
  n' \equiv n - |\kappa| \,,
\quad
  \gamma \equiv \sqrt{\kappa^2 - \alpha^2 Z_{n \kappa}^2} \,,
\quad
  N \equiv \sqrt{n^2 - 2 n' (|\kappa| - \gamma)}\,,
\quad
    \epsilon
\equiv 
    \frac{m_e + E_{n \kappa}(Z_{n \kappa})}{m_e}.
\label{paramters_bound}
\end{equation}
The corresponding energy eigenvalue $E_{n \kappa}$
is mainly a function of $N$ and the effective charge $Z_{n \kappa}$ \cite{Dyall2007IntroductionTR}, 
\begin{equation}
  E_{n, \kappa} (Z_{n \kappa})
=
  \frac{-Z_{n \kappa}^2}{N^2}
  \left[
1 + \sqrt{1 - 
\left(\frac{\alpha Z_{n \kappa}}{N} \right)^2}
  \right]^{-1}
  \times 27.2 \,{\rm eV},
\label{Binding_nk}
\end{equation}
with the fine-structure constant $\alpha$ denoting
the higher-order correction. The energy eigenvalue
ranges from $\mathcal O$(10)\,eV to $\mathcal O$(10)\,keV,
which is much smaller than the electron mass $m_e$,
and thus $\epsilon \sim 1$. In \geqn{Rel_Bound_Wave_L},
the key difference between the radial wave functions
$P$ and $Q$ lies in the factor $\sqrt{1 \pm \epsilon}$.
For those electrons in outer layers with a small energy
$E_{n \kappa}$, $1 -\epsilon \simeq 0$, which leads to
tiny values for the $Q$ function.
This is the reason why 
$P$ is usually called as the large component while 
$Q$ the small component of the radial wave functions.
Together, they satisfy a normalization condition, 
\begin{equation}
  \int
\left[
  |P_{n \kappa}(|\bm r|)|^2
+ |Q_{n \kappa} ( |\bm r|)|^2
\right] d |\bm r|
= 1,
\end{equation}
which is consistent with \geqn{eq:bound-norm}.

\subsubsection{Ionized Radial Wave Function}

For an electron in the ionized state $\ket{T_r \kappa m}$,
the corresponding radial wave function
\cite{Dyall2007IntroductionTR,2007rqta.book.....G} is,
\begin{align}
  P (Q)_{T_r \kappa}(|\bm r|)
& =
\mathcal{N}
\left[ \frac{(m_e + T_r) \pm m_e }{ (T_r+ m_e) \mp m_e } \right]^{1 / 4}
(-2 i |{\bm p}| |\bm r|)^\gamma 
\mathrm{e}^{i |{\bm p}| |\bm r|} 
\label{Rel_Ionized_L}
\\
& \times
\left[-\frac{\gamma-i \nu}{\kappa-i \nu^{\prime}} 
{}_1F_1(\gamma+1-i \nu, 2 \gamma+1,-2 i |{\bm p}| |\bm r|)
\pm
{}_1F_1(\gamma-i \nu, 2 \gamma+1,-2 i |{\bm p}| |\bm r|)\right],
\nonumber
\end{align}
where the asymptotic momentum 
$|{\bm p}|$ is defined as
$|{\bm p}| \equiv \sqrt{(m_e + T_r)^2 - m_e^2}$ and the
parameter $\gamma$ is defined in \geqn{paramters_bound}.
The other two new parameters are 
\begin{equation}
  \nu (Z_{n_i \kappa_i}, T_r)
\equiv
  \frac{Z_{n_i \kappa_i} (m_e + T_r)}{| {\bm p}| }
\quad \mbox{and} \quad
    \nu' (Z_{n_i \kappa_i}, T_r) \equiv \frac{Z_{n_i \kappa_i} m_e}{|{\bm p}|} .
\label{eq:nu}
\end{equation}
For an electron that is ionized from the initial state
$\ket{n_i,\kappa_i}$, it should share the same effective
charge $Z_{n_i \kappa_i}$ as the initial state which has
been implemented in the code of \cite{Catena:2019gfa}.
This is because the other electrons and the nuclei that
contribute the effective charge are not affected during
the process.

As shown again in \geqn{Rel_Ionized_L}, the small radial
component wave function $Q$ is highly suppressed
for $ T_r \ll m_e$.
The normalization condition \geqn{normFinal} for the
ionized radial wave function, or equivalently,
\begin{equation}
  \int_0^{\infty}
\left[
  P^*_{T_r \kappa}(|\bm r|) P_{T_r \kappa}(|\bm r|)
+ Q^*_{T_r \kappa}(|\bm r|) Q_{T_r \kappa}(|\bm r|)
\right] d |\bm r|
=
  2 \pi,
\label{norm-rel}
\end{equation}
requires that the normalization factor $\mathcal N$
in \geqn{Rel_Ionized_L} should be \cite{2007rqta.book.....G},
\begin{equation}
    \mathcal N
=
     \sqrt{2\pi}(\pi c)^{-1/2}
\left[
    \frac 2 {\gamma + i \nu}
    \frac{\Gamma (2 \gamma+1)}{{\rm Re}(\Gamma (\gamma - i \nu))}
    \sqrt{\frac{\gamma+i\nu}{-\kappa + i \nu'}}
    e^{-i \gamma \pi/2}
    e^{-\nu \pi/2}
\right]^{-1}.
\end{equation}

\subsubsection{The Inner Product of Radial Wave Functions}
\label{sec:Rel-K}

With the quantum state and wave function of
both the initial and final electrons defined, 
the inner product of states in \geqn{kernel3} can
be explicitly expanded as,
\begin{align}
\left\langle
  T_r \kappa' \left. \frac12 
  \right| \gamma^0_D T_{L 0}
  \left| n \kappa \frac 1 2 \right.
\right\rangle
& =
  4 \pi i^L
  \int d^3 \bm r  j_{L} (|\bm q| r) Y_{L0} (\hat{\bm r})
    \psi^\dagger_{T_r \kappa' \frac12}
    \gamma^0_D
    \psi_{n \kappa \frac12}
\nonumber
\\
& \hspace{-5cm} = 
4 \pi i^L
    \left(
    a a'
    \sqrt{\frac{l + \frac12 + \frac12 a }
     {2 l +1 }}
    \sqrt{\frac{l' + \frac12 + \frac12 a' }
     {2 l' +1 }}
    \mathcal{R}_{PP} \Omega_{l' , L, l}^0
    +
    \sqrt{\frac{l + \frac12 - \frac12 a }
     {2 l +1 }}
    \sqrt{\frac{l' + \frac12 - \frac12 a' }
     {2 l' +1 }}
    \mathcal{R}_{PP} \Omega_{l', L, l}^1
    \right.
\nonumber
\\
& 
    \hspace{-5cm} -
    \left.
    a a'
    \sqrt{\frac{l_s + \frac12 - \frac12 a }
     {2 l_s +1 }}
    \sqrt{\frac{l'_s + \frac12 - \frac12 a' }
     {2 l'_s +1 }}
    \mathcal{R}_{QQ} \Omega_{l'_s, L, l_s}^0
    -
    \sqrt{\frac{l_s + \frac12 + \frac12 a }
     {2 l_s +1 }}
    \sqrt{\frac{l'_s + \frac12 + \frac12 a' }
     {2 l'_s +1 }}
    \mathcal{R}_{QQ} \Omega_{l'_s, L, l_s}^1
    \right).
\label{Scalar_Inner}
\end{align}
The sign parameter $a (a')$ and the orbital angular
momentum $l_{(s)}$ ($l'_{(s)}$) are for the initial (final) state.
The integrations $\mathcal R_{PP}$ and 
$\mathcal R_{QQ}$ of the radial wave functions
are defined as, 
\begin{subequations}
\begin{align}
    \mathcal R_{PP}
& \equiv
    \int P_{T_r \kappa'}^* (|\bm r|) j_L(|\bm q| |\bm r|) P_{n \kappa} (|\bm r|) d |\bm r|
\\
\mathcal R_{QQ}
& \equiv
    \int Q_{T_r \kappa'}^* (|\bm r|) j_L(|\bm q| |\bm r|) Q_{n \kappa} (|\bm r|) d |\bm r|,
\end{align}
\label{radial_integral}
\end{subequations}
while the angular integration is, 
\begin{gather}
    \Omega_{l_1, l_2, l_3}^m
\equiv
    \int d \Omega_{\bm r} 
    Y^{m*}_{l_1} (\hat{\bm r})
    Y_{l_2}^0 (\hat{\bm r})
    Y_{l_3}^m (\hat{\bm r}).
\label{angular_int}
\end{gather}
After applying the Wigner–Eckart theorem,
the tensor operator has a magnetic quantum number
$M=0$ as shown in \geqn{kernel3}. With conservation
of the magnetic quantum number of orbital angular
momentum, the initial and final states must share
the same magnetic quantum number $m$.

We comment here that, by simply replacing the $\gamma$-matrix structure appearing in \geqn{Scalar_Inner}, the 
relativistic scattering cross section and
$K$-factor formalism can be generalized to other interactions. For example, in the pseudo-scalar interaction case, the $K$-factor is obtained by replacing the corresponding term in \geqn{kernel3} with the inner production,
\begin{align}
  \braket{  T_r \kappa' \frac12 
  | T_{L0} \gamma^0_D \gamma^5_D
  | n \kappa \frac12 }
& =
  4 \pi i^L \int d^3 \bm r j_{L} (|\bm q| r) Y_{L0} (\hat{\bm r})
  \psi^\dagger_{T_r \kappa' \frac12}
  \gamma^0_D \gamma^5_D
  \psi_{n \kappa \frac12}
\label{FinalInnerPPS}
\\
& \hspace{-4.5cm} = 
4 \pi i^L
    \left(
    -i a a'
    \sqrt{\frac{l_s + \frac12 - \frac12 a }
     {2 l_s +1 }}
     \sqrt{\frac{l' + \frac12 + \frac12 a' }
     {2 l' +1 }}
    \mathcal{R}_{PQ} \Omega_{l', L, l_s}^0
    +
    i 
    \sqrt{\frac{l_s + \frac12 + \frac12 a }
     {2 l_s +1 }}
    \sqrt{\frac{l' + \frac12 - \frac12 a' }
     {2 l' +1 }}
    \mathcal{R}_{PQ} \Omega_{l', L, l_s}^1
    \right.
\nonumber
\\
&   \hspace{-4.5cm}
    \left.
    -i a a'
    \sqrt{\frac{l + \frac12 + \frac12 a }
     {2 l +1 }}
    \sqrt{\frac{l'_s + \frac12 - \frac12 a' }
     {2 l'_s +1 }}
    \mathcal{R}_{QP} \Omega_{l'_s, L, l}^0
    +
    i 
    \sqrt{\frac{l + \frac12 - \frac12 a }
     {2 l +1 }}
    \sqrt{\frac{l'_s + \frac12 + \frac12 a' }
     {2 l'_s +1 }}
    \mathcal{R}_{QP} \Omega_{l'_s, L, l}^1
    \right) .
\nonumber
\end{align}
Meanwhile, the Lorentz structure in the DM spinor summation should be modified accordingly.

The vector (V) or axial-vector (A) cases are slightly more involved, but the basic feature should be the same. One needs to insert $\Gamma^{V(A)} = \gamma^\mu(\gamma^\mu \gamma^5)$ with $\mu = 1,2,3,4$ into the DM and electron bilinears, respectively, and then contract the two indices term by term. In this situation, a scalar atomic $K$-factor can no longer be defined. This is also where our approach differs from previous treatments that factorizes atomic effects through a single overall form factor \cite{Essig:2012yx,Chen:2015pha,Essig:2015cda,Roberts:2015lga,Roberts:2016xfw,Roberts:2016sem,Essig:2017kqs,Bloch:2020uzh,Chen:2021qao,Hamaide:2021hlp,Ge:2021snv,Emken:2021vmf}.

For the factorization approach \cite{Essig:2015cda} as summarized
in \geqn{previous_f}, the Lorentz structure is absorbed into
the free-electron scattering matrix element and the $K$-factor
itself carries no  $\gamma$-matrix structure.
Similarly, treating the DM and electron interaction
as an effective potential that is sandwiched
by the initial- and final-state electron wavefunctions
can also obtain a $K$-factor without Lorentz structure
\cite{Roberts:2015lga,Roberts:2016xfw,Roberts:2019chv}. To keep the Lorentz
structures, the electron field must be contracted directly
with the $\gamma$-matrices without assumption at the beginning.
It turns out that only for the scalar- and pseudoscalar-type interactions,
a factorizable $K$-factor is possible in the fully relativisitic
approach.

As can be seen from \geqn{Scalar_Inner}, the evaluation of the $K$-factor depends primarily on the inner product of the electron wave functions. Here, we obtain the wave functions using the Coulomb approximation with an effective charge. This framework can be generalized to eigenstates and wave functions computed with other many-body method \cite{Chen:2013iud,Chen:2016eab,Wu:2017zcd,Pandey:2018esq,Qiao:2020ybv},
such as Hartree–Fock \cite{Catena:2019gfa}
or density functional theory
\cite{Essig:2015cda,PhysRevD.104.095015,PhysRevD.109.115008},
without any modification to the form of the $K$-factor expression.

With a large momentum transfer $|{\bm q}|$ and 
angular momentum transfer $L$, both the final-state
wave function in \geqn{Rel_Ionized_L}
and the Bessel function $j_L (|{\bm q}| |\bm r|)$
are highly oscillatory. 
This makes the radial integrals and the summation
over $L$ in the $K$-factor numerically time-consuming
and challenging to compute
\cite{Catena:2019gfa,Ge:2021snv}. The numerical bottleneck can be alleviated by performing an analytical preprocessing of the radial wave functions and the associated radial integrals as summarized in \gapp{numerical}.

\subsection{Relativistic Effects in Radial Wave Functions}

\label{sec:relativisticEffects@Waves}

The non-relativistic limit is satisfied 
when the electron in scattering has a negligible kinetic  
energy. For the 
bound state, 
it means the binding energy, or 
equivalently, the effective charge $Z_{n \kappa}$,
is small according to the Virial theorem.  For
the ionized case, it means the recoil 
energy $T_r$ is tiny.
In both cases, the small component $Q$ of
the radial wave function becomes much smaller
than the larger one $P$, $Q/P \ll 1$.
As mentioned below \geqn{Binding_nk}, the
$Q / P \sim E_{n \kappa} / 2 m_e$ is $\sim 0.04$
for the innermost state $\ket{n= 1, \kappa = -1}$
with a binding energy $E_{1,-1} \simeq 40\,$keV.
Furthermore, the radial integration $\mathcal R_{QQ}$ is
a second-order small quantity in comparison with the
leading contribution $\mathcal R_{PP}$.
As a result, the small components of wave functions can be neglected at the beginning 
and the four-spinor solution \geqn{wavefunction_general}
reduces to a two-spinor one with only the upper large components,
\begin{equation}
  \psi_{n(T_r) \kappa m_j}
\rightarrow
  \frac 1 {|\bm r|}
  P_{n(T_r) \kappa} \left(
  a \sqrt{\frac{l+\frac{1}{2}+a m_j}{2 l + 1}} 
  Y_{l}^{m_j-1 / 2}
  \eta_+
+
  \sqrt{\frac{l+\frac{1}{2}-a m_j}{2 l + 1}} Y_{l}^{m_j+1 / 2}
  \eta_- \right).
\label{P_reduction}
\end{equation}
The two-component spinor $\eta$ is
$\eta_+ = (1, 0)^T$ for the spin magnetic quantum
number $m_s = +1/2$ while 
$\eta_- = (0, 1)^T$ for $m_s = -1/2$.

However, in the non-relativistic quantum mechanics, 
the spin wave function
is a direct product with its spatial counterpart
rather than a coupled one. In addition,
the spin magnetic number $m_s$ is another
good quantum number besides the energy level $n$, 
orbital angular momentum $l$, and orbital magnetic
quantum number $m_l$.
Thus, the spinor wave function
of state $\ket{n (T_r) l m_l m_s}$ is, 
\begin{equation}
  \phi_{n (T_r) l m_l m_s} (|\bm r|, \Omega_{\bm r})
=
  \frac 1 {|\bm r|} R_{n (T_r) l} (|\bm r|) Y_{l}^{m_l} (\Omega_{\bm r}) \eta_{m_s}. 
\label{NRwavefunction}
\end{equation}
The radial function $R_{n (T_r) l}$ 
is a solution of the Schrodinger equation
with a hydrogen-like potential 
for the bound \cite{cohen1977quantum}
(ionized \cite{Catena:2019gfa,Bethe:1957ncq}) states.

The non-relativistic radial wave function $R$ and 
the relativistic large component wave function $P$
have a strict correspondence. Although the
non-relativistic wave function \geqn{DiracSolution1}
is solved in the chiral representation,
it can be transformed to the Dirac representation,
\begin{align}
  \psi_D = \mathcal R_D \psi_{a, m_s}
\approx
\left\lgroup
\begin{matrix}
  f_{a, m_s}(\bm r) \\
  \frac {i \bm \sigma \cdot \bm \nabla}
        {2 m_e}
  f_{a, m_s}(\bm r)
\end{matrix}
\right\rgroup,
\quad \mbox{with} \quad
  \mathcal R_D
\equiv
  \frac 1 {\sqrt 2}
\left\lgroup
\begin{matrix}
  I & I \\
  I &-I
\end{matrix}
\right\rgroup,
\end{align}
by a similarity transformation $\mathcal R_D$.
We can see that $f_{a, m_s}(\bm r)$ corresponds to the
large component with $P$ in \geqn{wavefunction_general}. 
It seems legitimate to direct compare the radial
function $R$ for the non-relativistic case with its
relativistic counterpart $P$.

\begin{table}[h]
\centering
\setlength{\arrayrulewidth}{0.4pt}
\renewcommand{\arraystretch}{1.2}
\begin{tabular}{|c|c|c|c|c|c|}
\hline
\multicolumn{1}{|c|}{States} & \multicolumn{1}{c|}{$3s$} & \multicolumn{2}{c|}{$3p$} & \multicolumn{2}{c|}{$3d$} \\
\hline
\multirow{2}{*}{Non-Relativistic}  
  & \(\makebox[1.5cm][c]{\(l = 0\)}\) 
  & \multicolumn{2}{c|}{\(\makebox[1.5cm][c]{\(l = 1\)}\)} 
  & \multicolumn{2}{c|}{\(\makebox[1.5cm][c]{\(l = 2\)}\)} \\
  & \(\makebox[1.5cm][c]{\(N_e = 2\)}\) 
  & \multicolumn{2}{c|}{\(\makebox[1.5cm][c]{\(N_e = 6\)}\)} 
  & \multicolumn{2}{c|}{\(\makebox[1.5cm][c]{\(N_e = 10\)}\)} \\
\hline
\multirow{4}{*}{Relativistic} 
  & \(\makebox[1.5cm][c]{\(\kappa = -1\)}\) 
  & \(\makebox[1.5cm][c]{\(\kappa = 1\)}\) 
  & \(\makebox[1.5cm][c]{\(\kappa = -2\)}\) 
  & \(\makebox[1.5cm][c]{\(\kappa = 2\)}\) 
  & \(\makebox[1.5cm][c]{\(\kappa = -3\)}\) \\
  & \(\makebox[1.5cm][c]{\(j = \frac{1}{2}\)}\) 
  & \(\makebox[1.5cm][c]{\(j = \frac{1}{2}\)}\) 
  & \(\makebox[1.5cm][c]{\(j = \frac{3}{2}\)}\) 
  & \(\makebox[1.5cm][c]{\(j = \frac{3}{2}\)}\) 
  & \(\makebox[1.5cm][c]{\(j = \frac{5}{2}\)}\) \\
  & \(\makebox[1.5cm][c]{\(l = 0\)}\) 
  & \(\makebox[1.5cm][c]{\(l = 1\)}\) 
  & \(\makebox[1.5cm][c]{\(l = 1\)}\) 
  & \(\makebox[1.5cm][c]{\(l = 2\)}\) 
  & \(\makebox[1.5cm][c]{\(l = 2\)}\) \\
  & \(\makebox[1.5cm][c]{\(N_e = 2\)}\) 
  & \(\makebox[1.5cm][c]{\(N_e = 2\)}\) 
  & \(\makebox[1.5cm][c]{\(N_e = 4\)}\) 
  & \(\makebox[1.5cm][c]{\(N_e = 4\)}\) 
  & \(\makebox[1.5cm][c]{\(N_e = 6\)}\) \\
\hline
\end{tabular}
\caption{The correspondence between the non-relativistic
and relativistic angular
momentum quantum states for the principal
quantum number $n = 3$. The electron capacity number
of the corresponding quantum state is
$N_e = 2(2l+1)$ for the non-relativistic case and
$N_e = 2 |\kappa|$ for the relativistic one.
For $\kappa$, its size $|\kappa| = j + \frac 1 2$
is determined by $j$ while its sign indicates the
spin-orbital coupling, $j = l - \frac 1 2 \mbox{Sign}(\kappa)$.}
\label{tab:l&k}
\end{table}
The correspondence between the quantum numbers
for the relativistic and non-relativistic cases can be
clearly seen in the correspondence between the state with
$l=0$ in the non-relativistic case and the state with
$\kappa=-1$ in the relativistic case as shown in \gtab{tab:l&k}.
When $\kappa=-1$ and hence $a \equiv - \mbox{Sign}(\kappa) = 1$,
the total angular momentum $j = l + a/2 = 1/2$ consists
only of the electron spin since $l = 0$.
The two electrons in this state correspond to the two
spin magnetic quantum numbers $m_j = m_s=+1/2$ with
a wave function $\psi = \frac 1 {|\bm r|} P Y_0^0 \eta_+$
and $m_j = m_s = -1/2$ with a wave function
$\phi = \frac 1 {|\bm r|} P Y_0^0 \eta_{\pm}$,
respectively, as reduced from \geqn{P_reduction}.
The two square root coefficients
$\sqrt{(l + 1/2 \pm a m_j) / (2 l + 1)}$
all reduces to 1. Correspondingly, the two
spin components for the non-relativistic case
are $\frac 1 {|\bm r|} R Y^0_0 \eta_\pm$ according to
\geqn{NRwavefunction}. One
can see the correspondence between the non-relativistic
and relativistic electron radial wave functions, $R \sim P$.

\begin{figure}[t]
\centering
\includegraphics[width=8.1cm]{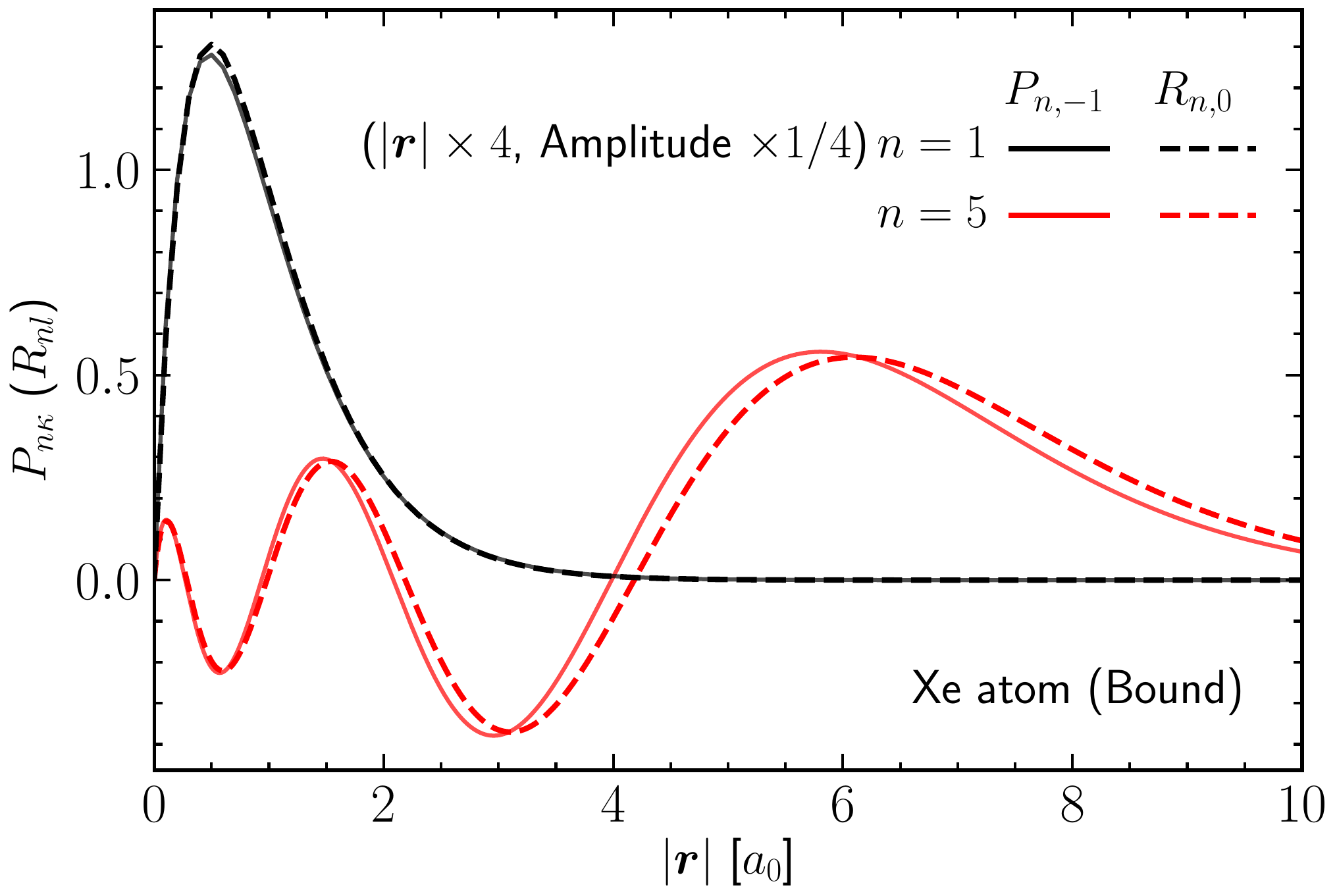}
\quad
\includegraphics[width=8.3cm]{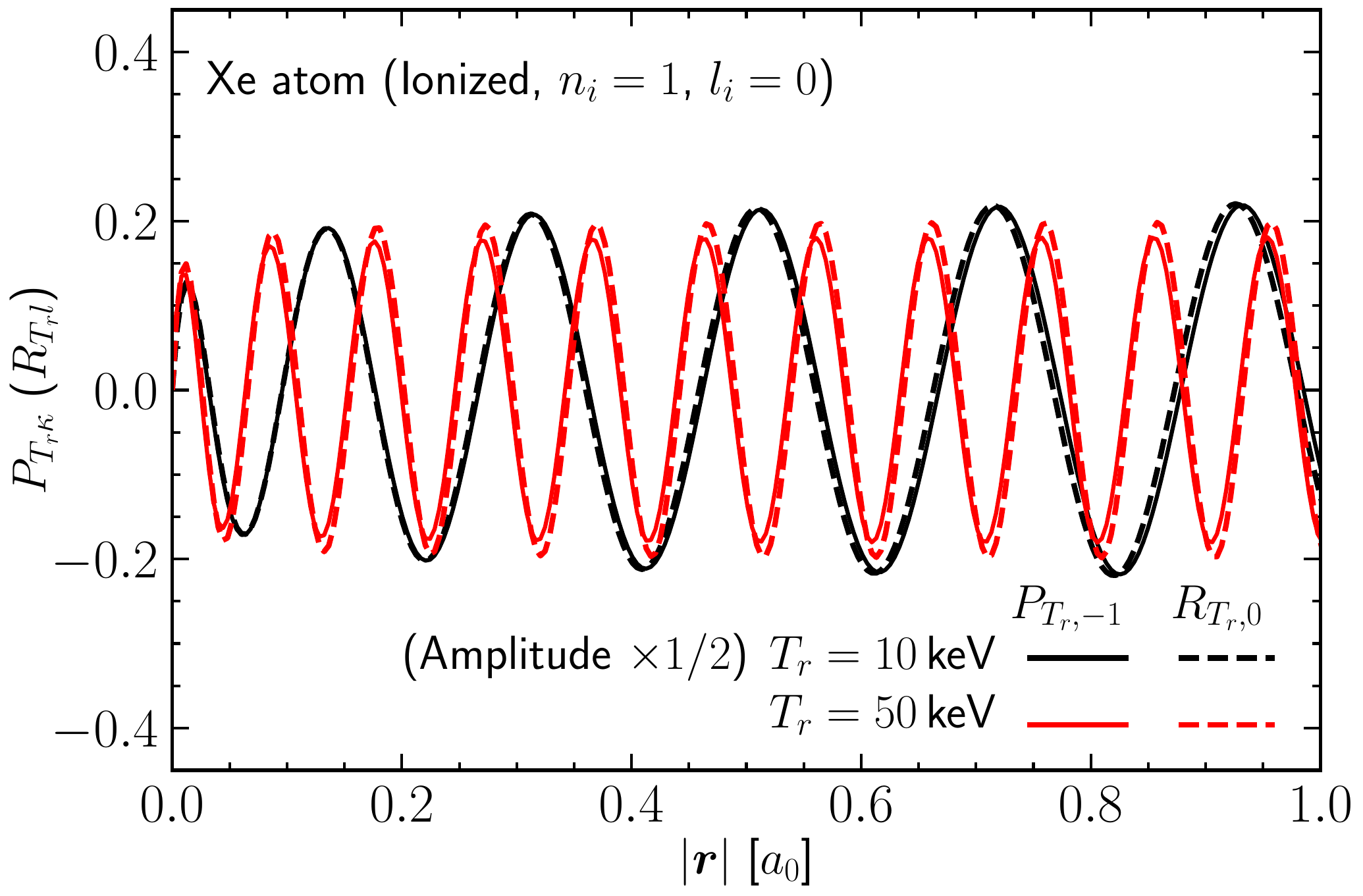}
\caption{
The bound (left panel) and ionized (right panel)
electron wave functions for typical quantum states.
The bound electrons are in the state $n = 1$ (black)
and $n = 5$ (red) with $\kappa = -1$ for the relativistic 
case (solid) and $l = 0$ for the non-relativistic
case (dashed) as shown in the left panel. 
The ionized states recoiled from initial state 
$n_i = 1$ and $l_i =0$
with recoiled energies $T_r = 10\,$keV (black) and
$T_r = 24\,$keV (red) are shown in the right panel.
Similarly, the dashed lines are for non-relativistic
wave functions and solid lines for relativistic
counterparts. To better display in the same plot,
some wave functions have been rescaled.
}
\label{waveFunctions}
\end{figure}

Since the kinetic energy dependence of the
$K$-factor enters through wave functions, 
we first make comparison at the wave-function
level as shown in \gfig{waveFunctions} while
the $K$-factor level comparison will be discussed
in the following \gsec{sec:orbit-spin}.
Both the relativistic (solid) and non-relativistic
(dashed) wave functions of bound electrons at the
energy levels $n = 1$ (red, rescaled by a factor of $1/4$)
and $n = 5$ (black) with orbital 
angular momentum $l = 0$ are shown in the left panel.
The solid and dashed lines exhibit a high degree
of similarity in terms of shape and amplitude.
The slight deviation between them stems from
the different binding energies and hence different
effective charges.
For example, the energy eigenvalue is
$E_{n\kappa} = 23.3\,$eV with an effective charge
$Z_{n \kappa} = 6.87$
for the relativistic state $\ket{n = 5,\kappa = -1}$
and $E_{nl} = 22.9\,$eV with an effective charge
$Z_{n l} =6.5$ for the non-relativistic state
$\ket{n = 5, l = 0}$ \cite{webelements_xenon}.
Since the Dirac equation contains 
higher-order perturbative expansion terms
that are not considered in the Schrödinger
equation with a central Coulumb potential in
\geqn{eq:Schrodinger}, the electron energy eigenvalues
are slightly different between the two cases. 
For example, the
kinetic term in non-relativistic quantum mechanics,
$T_r = |{\bm p}^2|/2m_e$ only retains the leading
term of the relativistic dispersion relation,
$T_r = \sqrt{|{\bm p}^2 + m_e^2|} - m_e$
\cite{cohen1977quantum,Sakurai:2011zz}.

The relativistic (solid) and non-relativistic (dashed)
wave functions of the ionized states with recoil energies 
$T_r = 10\,$keV (black) and $T_r = 50\,$keV (red) are shown
in the right panel of \gfig{waveFunctions}.
It is assumed that the electrons 
are ionized from the initial state $n_i = 1$, $l_i = 0$
($\kappa_i = -1$) with the corresponding effective charges. As expected, the relativistic effect
increases with the recoil energy. 
The black dashed and solid lines with
$T_r \simeq 2\% m_e \simeq 10\,$keV nearly
overlap with each other. On the other hand, 
the red dashed and solid lines with
$T_r \simeq 10\% m_e \simeq 50\,$keV exhibit
more apparent difference in amplitude. 
Take the second trough of the red lines at $|\bm r| \simeq 0.13\,a_0$, where $a_0 = 1/m_e \alpha$ is the Bohr radius, as an example, 
the wave function amplitude is $-0.197$ for the relativistic case and becomes $-0.174$ for the non-relativistic case. 
This discrepancy is around $13\%$.

\begin{figure}[t]
\centering
\includegraphics[width=8.1cm]{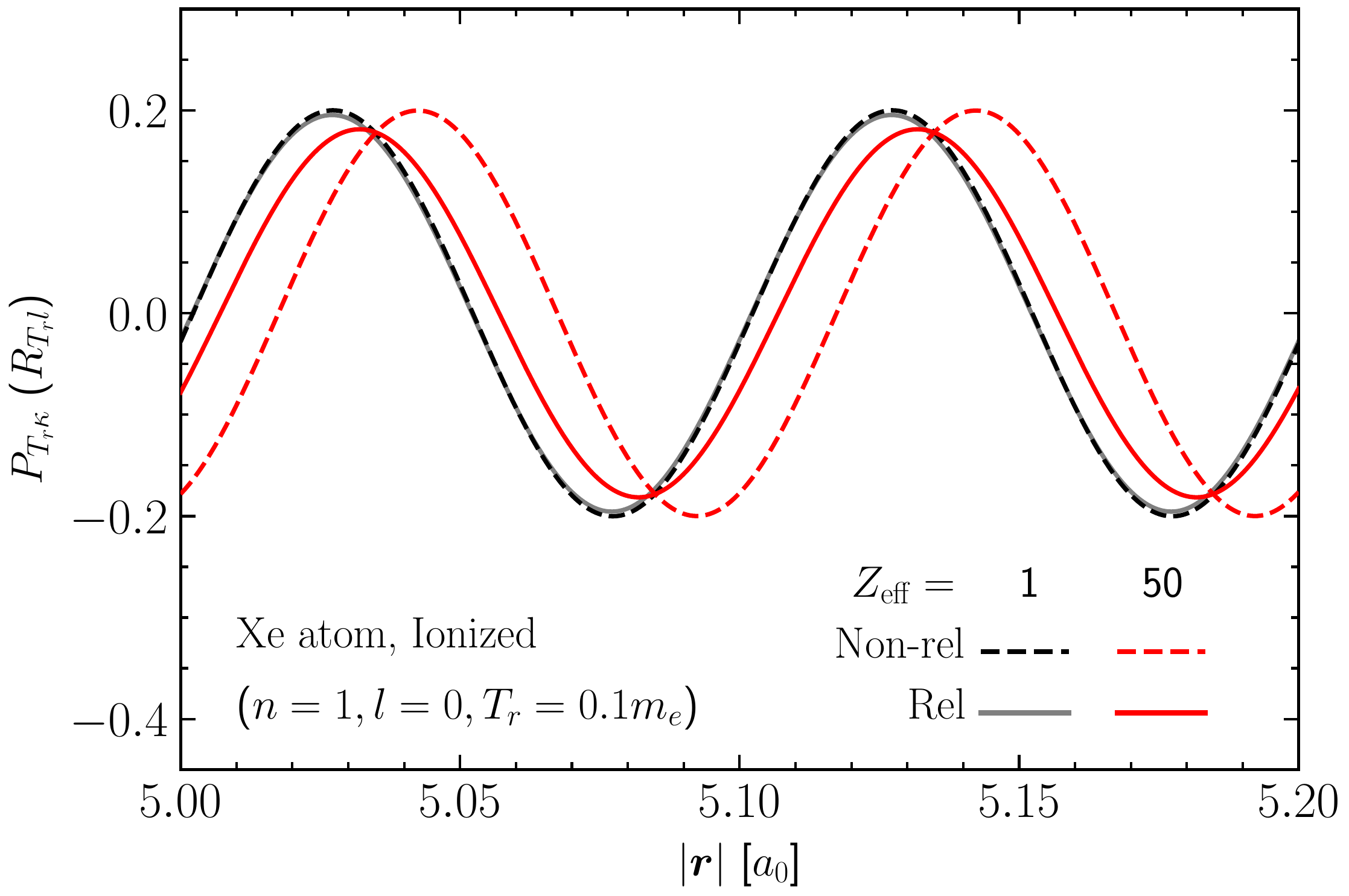}
\quad
\includegraphics[width=8.3cm]{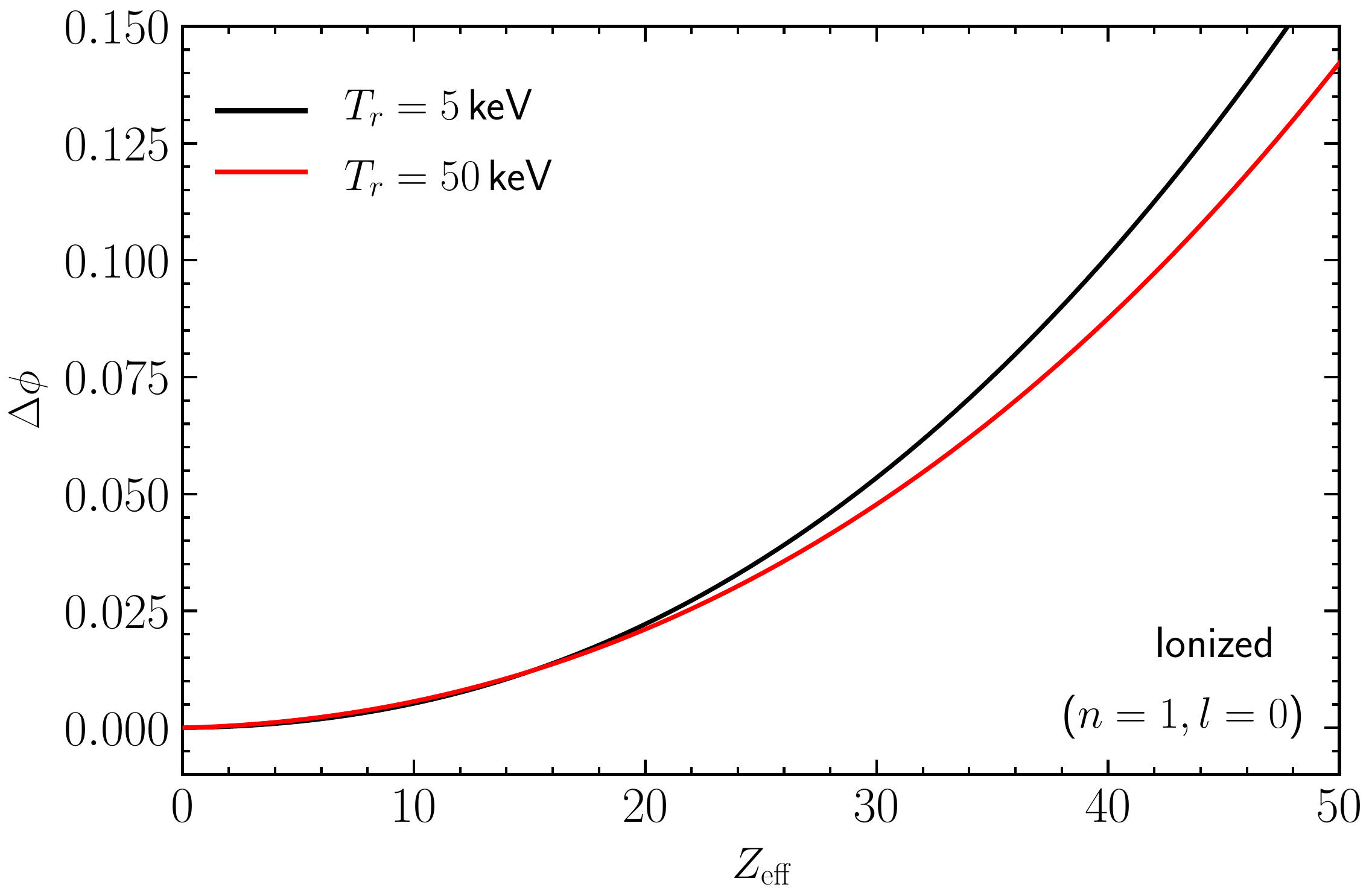}
\caption{
(Left) Ionized electron wave functions with the same quantum
numbers ($n = 1$, $l = 0$) and recoil energy ($T_r = 0.1\,m_e$) 
but different effective charges $Z_{\rm{eff}} = 1$ (black)
and $Z_{\rm{eff}} = 50$ (red). Both the non-relativistic (dashed)
wave function $R$ and the relativistic (solid) one
$P$ are shown for comparison.
(Right) The relationship between the phase difference
$\Delta \phi$ of the relativistic and non-relativistic
ionization wave functions and the effective charge
$Z_{\rm eff}$. The quantum numbers are chosen
as $n = 1$ and $l = 0$ while the black and red curves correspond
to recoil energies of $5$ and $50\,$keV respectively.
}
\label{phases}
\end{figure}

In addition to the difference in the amplitude that
depends on the recoil energy $T_r$, the main difference
between the relativistic and non-relativistic ionized
wave functions also lies in their relative phase difference.
At places with a large radius, the nuclear Coulomb
potential becomes smaller and the electron wave function
behavior approaches a plane wave. To make this feature
explicit, we take the asymptotic form
\cite{Bethe:1957ncq,2007rqta.book.....G},
\begin{subequations}
\begin{align}
& R_{T_r, l} \sim \cos 
\left[ |\bm p| |\bm r| + \frac{Z_{\rm{eff}} \, m_e}{|\bm p|} \ln (2 |\bm p| |\bm r|) - \frac{\pi}{2} (l+1) - \arg \Gamma \left(l+1 + i \frac{Z_{\rm{eff}} \, m_e}{|\bm p|}\right) \right],
    \label{eq:nonrel_asy}
    \\
& P_{T_r,l} 
    \sim \cos 
\left[ |\bm p| |\bm r| + \nu \ln (2 |\bm p| |\bm r|) - \frac{\gamma \pi}{2} - \arg \Gamma \left(\gamma + i \nu \right) + \frac{1}{2i} \ln \left( \frac{-\kappa + i \nu'}{\gamma+i \nu}\right)
\right],
\label{eq:rel_asy}
\end{align}
\end{subequations}
for the relativistic and non-relativistic wave functions. 
Here, $|{\bm p}| = \sqrt{(m_e+T_r)^2 - m_e^2}$
is the norminal (asymptotic) momentum defined in terms of
the recoil energy $T_r$ and it becomes
$|\bm p| \simeq \sqrt{2 m_e T_r}$ in the non-relativistic limit.
The other parameters $\gamma$, $\nu$, and $\nu'$ has
been previously defined in \geqn{paramters_bound} and
\geqn{eq:nu}, respectively.
Except for the $|\bm r|$-dependent part of the
first two terms, the constant terms are the phase shifts
caused by the nuclear Coulomb potential after scattering.
This phase shift is highly dependent on the effective charge
$Z_{\rm{eff}}$. With quantum numbers $n = 1$, $l = 0$,
and $T_r = 0.1\,m_e$, the left panel of \gfig{phases}
displays the ionized wave functions of relativistic
(solid) and non-relativistic (dashed) electrons when
$Z_{\rm{eff}} = 1$ (black) and $Z_{\rm{eff}} = 50$ (red).
It can be observed that when $Z_{\rm{eff}}$ is small,
the two wave functions almost coincide. As $Z_{\rm{eff}}$
increases to $50$, besides the amplitude difference discussed earlier, a significant phase difference between them becomes evident.

The phase difference $\Delta \phi$ is defined as
the distinction in phase shifts between the
relativistic and non-relativistic asymptotic
wave functions. Its dependence on the effective
charge $Z_{\rm{eff}}$ is illustrated in the
right panel of \gfig{phases}. As the effective
charge increases, the influence of the Coulomb
potential and relativistic effects becomes greater
such that the phase difference $\Delta \phi$
increases with $Z_{\rm{eff}}$. However, its
dependence on the recoil energy $T_r$ is not
significant, as the black and red lines
corresponding to $T_r = 5\,$keV and $T_r = 50\,$keV
show little difference.

When the orbital angular momentum $l$ becomes nonzero
($l \neq 0$), there would be no longer one-to-one
correspondence with the relativistic quantum number
$\kappa$. In \gtab{tab:l&k}, we list the $3p$
and $3d$ orbitals in the Xenon atom for illustration.
A non-zero $l$ can couple with spin in two ways to
form different total angular momenta as $j_1 = l - 1/2$
and $j_2 = l+1/2$. Therefore, a single $l$ corresponds
to two $\kappa$ numbers. However, the total number of
electrons with the same orbital angular momentum $l$
remains the same. In the non-relativistic case, the electron
capacity number is $N_e = 2 \times (2 l +1) = 4l+2$
considering the spin degeneracy. For comparison,
it is also $N_e = (2j_1 + 1) + (2 j_2 +1) = 4 l+2$
for the relativistic case.

%\subsection{\gred{Comparison Between Relativistic and Non-relativistic $K$-factors}}
\subsection{Relativistic Effects in $K$-Factor}
\label{sec:orbit-spin}

By substituting the corresponding wave functions and using our novel reduction strategy
of the radial integral as summarized in \gsec{sec:Rel-K}, the relativistic $K$-factors can now be calculated
to include the atomic effects into the scattering process. 
In this subsection, we will directly compare the $K$-factors in the relativistic and non-relativistic calculations to see the influence of relativistic effects on scattering.

As discussed in the last \gsec{sec:relativisticEffects@Waves},
the summation over the magnetic quantum numbers $m_j$
and $m'_j$ for the relativistic $K$-factor in \geqn{eq:K0}
actually contains the spin degrees of freedom.
However, the summation over $m_l$ and $m'_l$ in
the non-relativistic $K$-factor \geqn{K_nr}
contains only the orbital degrees of freedom.
To make an explicit comparison at the same
level of degrees of freedom between the relativistic
and non-relativistic cases, the non-relativistic
$K$-factor should also incorporate
the full spinor wave function
$u(m_e) \phi_{n(T_r)lm} /\sqrt{2m_e}$ with
$1/\sqrt{2m_e}$ being the extra normalization factor
\cite{Weinberg:1995mt,Ge:2021snv} and average over spins,
\begin{subequations}
\begin{align}
    \text{Non-rel:} &\quad  
    K_{nl} \equiv 
    \frac{\tr\left[\bar u (m_e) u(m_e) \bar u (m_e) u(m_e)\right]}{2(2l+1)(2m_e)^2}
    \sum_{l' m_{l'} m_l} 
    \left|
    \int d^3 \bm r
  e^{i\bm q \cdot \bm r}
  \phi_{T_r l' m'_l}^{*} (\bm r) 
  \phi_{n l m_l} (\bm r)
    \right|^2,
    \label{eq:nonrel}
    \\
    \text{Rel:} &\quad K_{n l}
\equiv
  \frac{1}{2|\kappa_1| + 2 |\kappa_2|}
    \sum_{\kappa_1, \kappa_2}\sum_{\kappa' m'_j m_j}
\left|
  \int d^3 {\bm r}
    e^{i {\bm q \cdot \bm r} }
    \psi^\dagger_{T_r \kappa' m'_j} ({\bm r}) 
    \gamma^0
    \psi_{n \kappa m_j} ({\bm r})
\right|^2.
\label{eq:rel}
\end{align}
\end{subequations}
To make things clear, the non-relativistic
$K$-factor above is defined for a single initial electron
but summed over all possible final states.
With spin summation, $u(m_e) \bar u(m_e)$ reduces to
$m_e (\gamma_0 + 1)$. Then the spinor trace reduces to
$\tr[m^2_e (\gamma_0 + 1)(\gamma_0 + 1)] = 8 m^2_e$ which
compensates the $8 m^2_e$ in the denominator where one
factor $2$ comes from the initial electron spin average.
Then \geqn{eq:nonrel} reduces to \geqn{K_nr}.

For the relativistic case, \geqn{eq:rel} being
normalized by $2 (|\kappa_1| + |\kappa_2|)$ seems to be quite
different from \geqn{kernel3} with normalization $2 \kappa$.
This is because the same initial orbital angular momentum
$l$ corresponds to two different initial quantum numbers
$\kappa$s, which we define as $\kappa_1$ and $\kappa_2$,
with the only exception of $l = 0$ according to \gtab{tab:l&k}.
To make exact comparison with the same orbital angular
momentum $l$, the relativistic $K$-factor also needs
to sum over the both quantum number $\kappa$s and
perform the corresponding averaging.
Note that the factor of 2 in normalization comes
from the initial electron spin average in the same way
as the non-relativistic case. So the relativistic
$K$-factor above is also for a single initial electron
and covers all possible final states.

For a clear comparison, we still start with the simplest
quantum state $l=0$ or equivalently
$\kappa = -1$ as summarized in \gtab{tab:l&k}.
Analogous to the procedure outlined in \gapp{Wigner_theorem},
the exponential term in \geqn{eq:nonrel} can also be
expanded as spherical harmonics $Y^*_{LM} (\hat{\bm q})$
and $Y_{LM} (\hat{\bm r})$. Additionally, by applying
the Wigner-Eckart theorem, the non-relativistic
$K$-factor decomposes into separable integrals over
the radial wave functions
$\int d |\bm r| R_{T_r l'}^* j_L(|{\bm q}| |\bm r|) R_{nl} (|\bm r|)$
and spherical harmonics $\Omega^0_{l' L l}$
as defined in \geqn{angular_int}. 
Then the non-relativistic $K$-factor in \geqn{eq:nonrel}
reduces to \geqn{eq:nonrel_simp},
\begin{subequations}
\begin{align}
    \text{Non-rel:} &\quad  
    K_{n,l = 0} (\Delta E, {\bm q})
=
    \sum_{l'=L} \sum_{L}
    4 \pi (2 L + 1) \left|
   \int dr R_{T_r l'}^* j_L(|{\bm q}| |\bm r|) R_{nl} (|\bm r|) 
    \right|^2 
    \left|
    \Omega^0_{l' L l}
    \right|^2,
    \label{eq:nonrel_simp}
    \\
    \text{Rel:} &\quad K_{n,\kappa = -1} (\Delta E, {\bm q})
 = 
    \sum_{\kappa'}
    \frac{l' + \frac 1 2 + \frac 1 2 a'}{2 l' +1}
    \sum_L 4 \pi (2 L + 1)
\nonumber
\\
     & \hspace{33mm} \times 
    \left|
  \int dr P_{T_r \kappa'}^* j_L(|{\bm q}| |\bm r|) P_{n\kappa} (|\bm r|) 
    \right|^2  \left|
    \Omega^0_{l' L l}
    \right|^2,
\label{eq:rel_simp}
\end{align}
\end{subequations}
with more comprehensive derivations to be found
in \cite{Ge:2021snv}. Since the initial state
has a magnetic qunatum number $m_l = 0$, the
requirement of \geqn{angular_int} automatically
enforces the final state to have $m'_l=0$ as well.

For the relativistic case, \geqn{kernel3} is the form
of \geqn{eq:rel} after being simplified by the Wigner-Eckart theorem.
Substituting the quantum number $\kappa = -1$ yields
the normalization factor $1/2$ while the Wigner-3j
symbol provides an additional factor
$((j',L,j), (- \frac 1 2,0, \frac 1 2) )^{-2} = 2 (2L+1)$.
Subsequently, the inner product of wave functions
in \geqn{kernel3} can be substituted by \geqn{Scalar_Inner}
to give \geqn{eq:rel_simp}.
Here, the second-order small quantity
$\mathcal R_{QQ}$ has been neglected.
Furthermore, with the quantum numbers $\kappa = 1$,
$l = 0$, $j = 1/2$, and $a = 1$, the factor
$\sqrt{l + 1/2 - a/2}$ vanishes and thus the
second term in \geqn{Scalar_Inner} also disappears,
and only the first term of \geqn{Scalar_Inner}
makes the major contribution.

The simplified $K$-factors in \geqn{eq:nonrel_simp}
and \geqn{eq:rel_simp} have the same prefactor
$4 \pi (2L+1)$ and angular integration $\left|
    \Omega^0_{l' L l}
    \right|^2$.
Furthermore, due to the correspondence of wave functions
$R \sim P$, their radial integrals with Bessel functions
are also similar.
Comparison between the non-relativistic $K$-factor \geqn{eq:nonrel_simp}
with the relativistic $K$-factor \geqn{eq:rel_simp}
shows their difference lies in the extra factor
$(l' + 1/2 + a'/2)/(2 l' +1)$ for the relativistic case.
Such a factor is always smaller than $1$ 
and approaches to $1/2$ with a large $l'$.  
However, the condition of angular momentum conservation
selects $l'$ as $l' = L$ with $l = 0$
as coded in $\Omega^0_{l' L l}$. A fixed $l'$ 
corresponds to two different $\kappa'$ as summarized in \gtab{tab:l&k}.
It is interesting to see that the radial wave function
$P_{T_r \kappa'}$ ($P_{n \kappa}$) of these two different
$\kappa'$ ($\kappa$) that corresponds to the same $l'$
($l$) are quite similar, for both the ionized and bound states.
Then the radial integration of \geqn{eq:rel_simp} becomes
essentially the same for the two final-state $\kappa'$.
Consequently, the summation over these two different $\kappa'$s
gives an additional factor of $2$ to cancel out the above $1/2$.
Putting things together,
$\sum_{\kappa'} (l' + 1/2 +  a'/2) / (2 l' + 1)$
effectively reduces to $\sum_{l' = L}$.
Therefore, the $K$-factors in the relativistic and non-relativistic
cases still have the same structure although the summation form differs.

This similarity can also be understood from the physical picture below.
To ionize a fixed initial state $|i \rangle$ to an ionized
state with energy $T_r$, the $K$-factor includes a summation
over all different angular momentum transfer operators $Y_L^0$
as explained below \geqn{angular_int}.
In the non-relativistic case, both the orbital angular momentum
and spin are good quantum numbers. And the orbital and
spin degrees of freedom can be detached from each other.
Then, the transferred angular momentum 
$L$ all goes to the orbital angular momentum as, 
$l + L \rightarrow l'$, without changing the spin
of electrons. Taking the initial state
$\ket{i} = \ket{l , m_l = 0} \eta_+$ as an
illustration, it can transform into final states
$\ket{f} = \ket{l' \subset (L-l, L +l), m_l = 0} \eta_+$
after being acted upon by the operator $Y^0_L$
that does not alter the magnetic quantum
number, namely $m'_l = m_l = 0$. Then, the
transition probability in the angular part $T_A$
is the summation over the final-state quantum
numbers of the inner products,
\begin{equation}
  T_A 
\equiv
  \sum_{l', m'_s} 
\left|
  \eta^\dagger_{m'_s} \bra{l', m'_l = 0}  Y_L^0 \ket{l, m_l = 0} \eta_+
\right|^2
=
  \sum_{L-l \leq l' \leq L+l} |\Omega_{l', L, l}^0|^2,
\end{equation}
with $m'_s$ being fixed to $1/2$ by $\eta_+$.
Since spin is a good quantum number in the non-relativistic
case, the summation over the final-state spins can only
select the part where the spin remains unchanged,
$m'_s = m_s$.

On the other hand, the only good quantum number in the
relativistic case is $\kappa$ or equivalently the total angular
momentum $j$,
which is a result of orbit-spin coupling. For a fixed
initial electron with a non-zero orbital angular momentum $l$,
it can have two different $\kappa$ quantum numbers
as summarized in \gtab{tab:l&k}. On the other hand,
a $\kappa$ only corresponds to a unique $l$. Consequently,
if a specific $\kappa$ is chosen,
the initial electron state $\ket i = \ket{\kappa, m_j = 1/2} =
    C_1 \ket{l, m_l = 0} \ket{\eta_+} 
    +C_2 \ket{l, m_l = 1} \ket{\eta_-}$
becomes a linear combination of two states with the same $l$.
To match the same total magnetic quantum number $m_j = 1/2$,
the orbital one $m_l$ should be either $0$ with a spin-up
$\eta_+$ or $1$ with a spin-down $\eta_-$.
The Clebsch–Gordan coefficients should satisfy the normalization condition $|C_1|^2 + |C_2|^2 = 1$. Similarly, after receiving an orbital angular momentum transfer $Y^0_L$, it can transfer to any final state containing the orbital angular momentum $l' \subset (L-l, L+l)$ 
as $\ket{f} = \ket{\kappa', m'_j = \frac 1 2} = \ket{l', m'_l = 0} \ket{\eta_+} /\sqrt{2}
+\ket{l', m'_l= 1} \ket{\eta_-}/\sqrt{2}$.
For a large $L$ and hence large $l'$ which is usually
the case when doing the summation in the $K$-factors, 
these two superposition coefficients are approximately $1/\sqrt{2}$.
The transition probability of angular parts is then, 
\begin{equation}
\begin{split}
    T_A 
& = \sum_{\kappa'(L-l \leq l' \leq L+l)} 
\left[ \left(\frac{1}{\sqrt{2}} \bra{\eta_+}\bra{l', m'_l = 0}
+
\frac{1}{\sqrt{2}} \bra{\eta_-}\bra{l', m'_l = 1}
\right) Y_L^0 \ket{i} \right]^2 \\
& = 
2 \sum_{L-l \leq l' \leq L+l} 
\left[ \frac{C_1^2}{2}  \left(
\Omega^0_{l',L,l}
\right)^2 
+ \frac{C_2^2}{2}
\left(
 \Omega^1_{l',L,l}
\right)^2 \right].
\end{split}
\end{equation}
Again, the factor of $2$ comes from the degeneracy of $\kappa'$
for a single $l'$.
Although both the initial $|i \rangle$ and final
$|f \rangle$ states have two terms, their inner product
has only two (instead of four) contributions in the second line of the above
equation. This is because the orbital angular momentum addition
with $Y^0_L$ requires $m_l$ and $m'_l$ to be the same,
$m_l = m'_l$ could be either 0 or 1 which lead to
$\Omega^0_{l',L,l}$ and $\Omega^0_{l',L,l}$, respectively.
The approximation $\Omega^1_{l',L,l} \simeq \Omega^0_{l',L,l}$
also holds in the large $L$ limit.
Therefore, $T_A \simeq \sum_{L-l \leq l' \leq L+l} 
(\Omega^0_{l',L,l})^2$ approaches the non-relativistic case.

However, there is still sizable difference between
the non-relativistic and relativistic results.
Before showing the concrete values, we first define
a proper variable for quantitative comparison.
According to our precious work \cite{Ge:2021snv},
the \textit{phase space ratio} $\mathcal R(T_r)$
is a good measure of the atomic effects. The differential 
cross section with respect to the recoil energy $T_r$
is an integration of all the contributions from different 
momentum transfers $\bm q$.
In the free electron scattering case, the initial electron 
is assumed to be stationary, where the momentum transfer
$|{\bm q}|$ is equivalent to the final-state electron
momentum $|{\bm p}'|$. Its phase space integration
is simply $\int |{\bm q}|^2 \delta (|{\bm q}| - |{\bm p}'|) d|{\bm q}| = |{\bm p}'|^2 = (m_e + T_r)^2 - m_e^2 \simeq 2m_e T_r$. 
In the case of bound electron scattering, the $K$-factor with a suitable normalization according to the final wave function describes 
the contributions of different momentum transfers for a scattering process. 
As a result, its phase space integration is
$\int |{\bm q}| K (\Delta E, {\bm q}) d|{\bm q}|$ which can be seen in \geqn{dcsWithK}.
The phase space ratio $\mathcal{R}$ defined as $\mathcal{R} (T_r) \equiv \int |{\bm q}| K (\Delta E, {\bm q}) d|{\bm q}|/2 m_e T_r$
measures how the atomic effects affect the 
differential cross section of DM-electron scattering.
This is particularly accurate in the heavy-mediator
scenario since the propagator $D^2(q)$ in \geqn{dcsWithK}
is independent of momentum transfer and nearly all
the $|{\bm q}|$-dependence in the integral comes from the $K$-factor.

\begin{figure}[t]
\centering
\includegraphics[width=8.2cm]{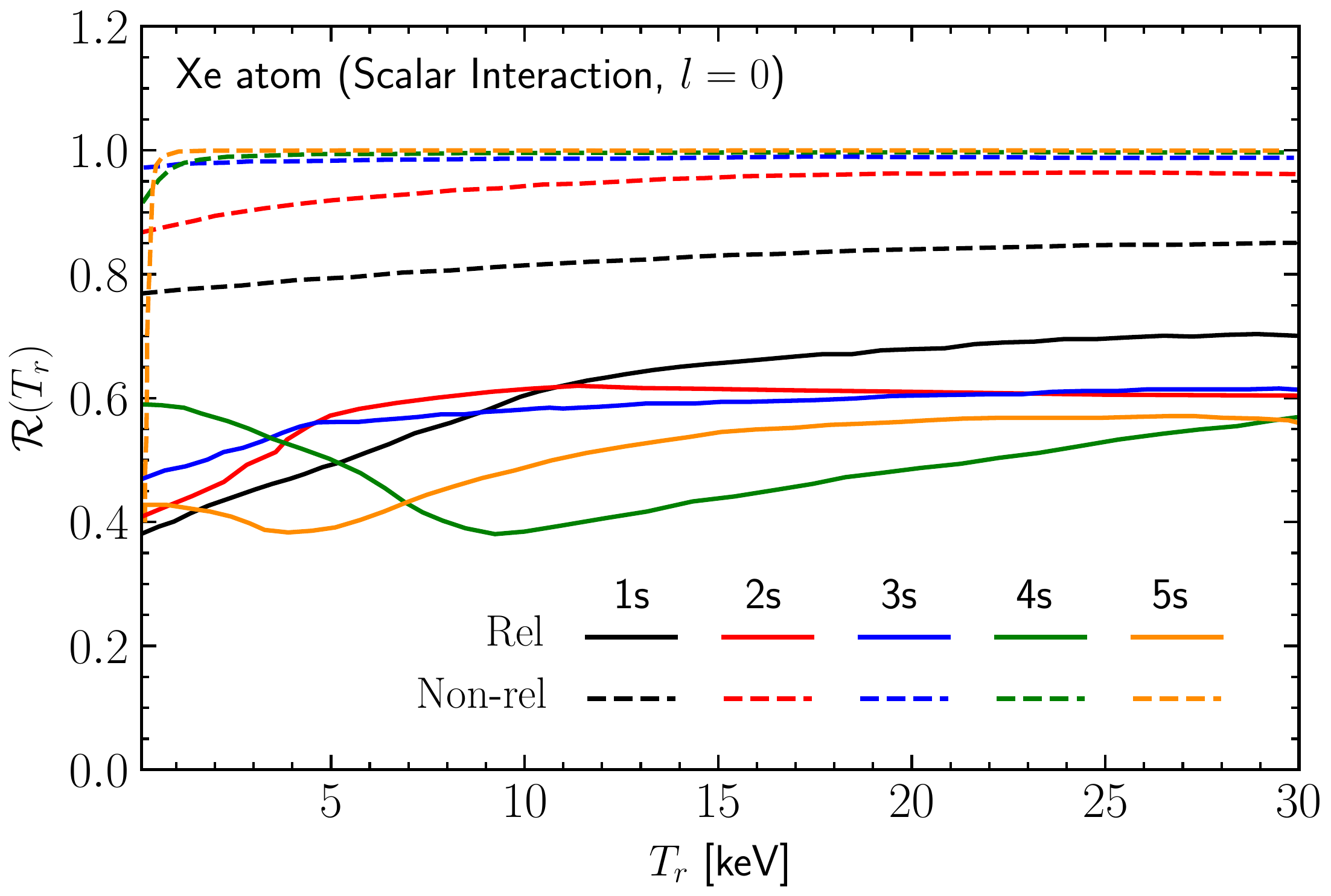}
\quad
\includegraphics[width=8.3cm]{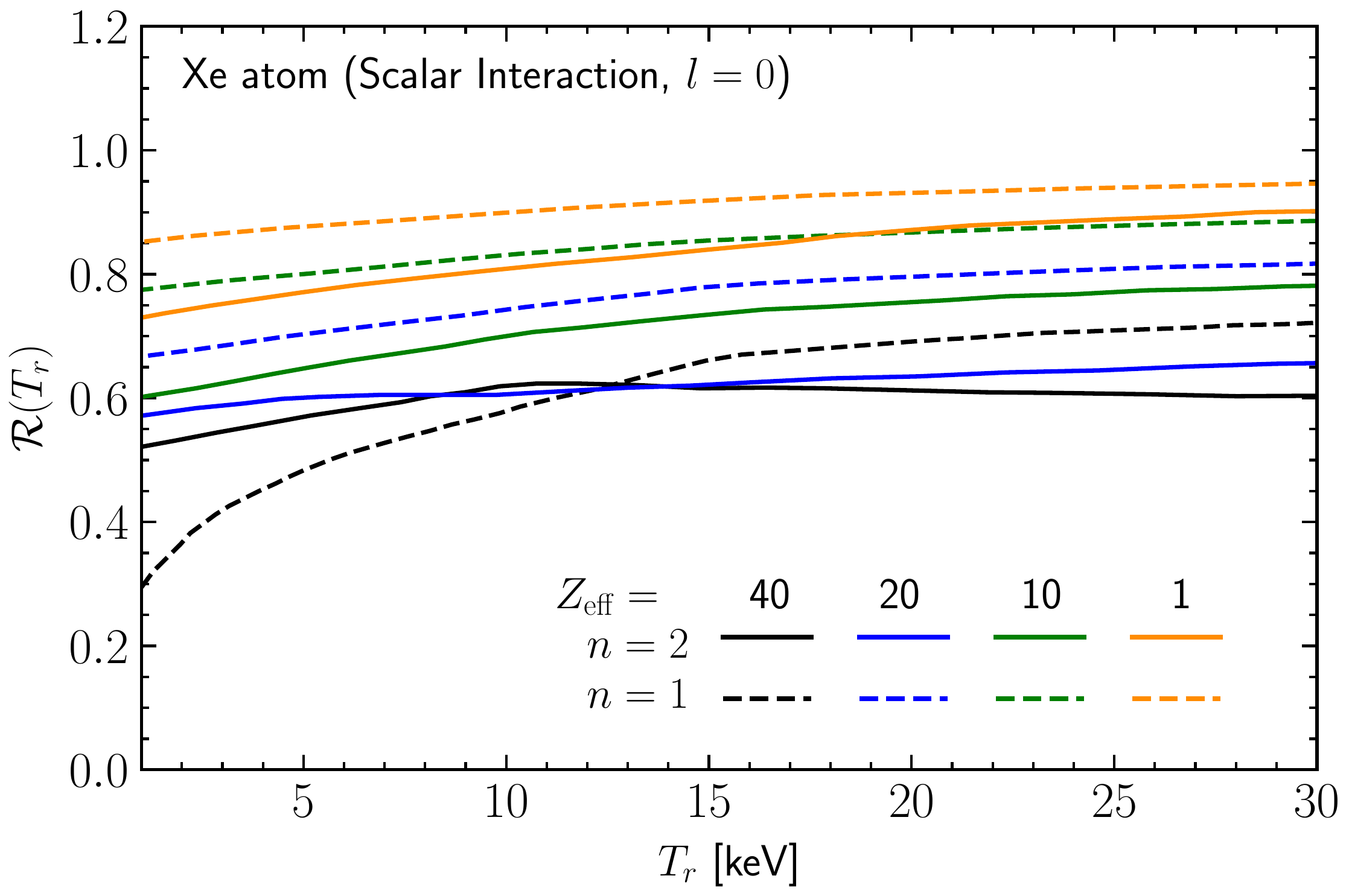}
\caption{
The comparison of the phase space ratios $\mathcal{R}(T_r)$ between the
non-relativistic (NR, dashed) and relativistic (Rel, solid) $K-$factors. The states with same angular quantum number 
$l = 0$ (or equivalently, $\kappa = -1$) but different 
principle quantum numbers $n = 1 \sim 5$ are shown with different colors in the left panel. The states with same 
principle quantum number $n = 3$ but different angular momentum $l = 0 \sim 3$ are shown in the right panel.
}
\label{fig:PhaseSpaceRatio_v2}
\end{figure}

The non-relativistic phase space ratios for the
$s$ sub-shell electrons ($l = 0$) with energy 
levels $n = 1 \sim 5$ are shown as dashed 
lines with different colors in the left panel of 
\gfig{fig:PhaseSpaceRatio_v2}. As the recoiled 
energy $T_r$ increases, all these ratios tend to 
approach $1$.
This means the atomic effects become 
negligible and the bound-electron cross sections
go back to the free case \cite{Ge:2021snv}. 
In comparison, the phase space ratios of 
the corresponding states with
$n = 1\sim 5$ and $\kappa = -1$ in the 
relativistic case, shown as solid lines in various colors,
have decreased around $30\% \sim 50\%$
compared to their non-relativistic counterparts.
This reduction in $K$-factor can be understood from 
the decrease of the final-state wave function
amplitudes shown in \gfig{waveFunctions}
and their phase shifts in \gfig{phases}.
The $K$-factor and the corresponding
suppression of the phase space ratio discussed
are derived using the scalar-type
interaction. However, they can also be extended
to other interaction types. Our analysis
indicates that this reduction is driven by changes
at the level of wave functions rather than by the
analytic expression of the $K$-factor itself.
Therefore, similar relativistic effects are
expected to appear for other types of interactions as well.

The suppression effect will decrease with
$Z_{\rm{eff}}$. The right panel of
\gfig{fig:PhaseSpaceRatio_v2} shows the
dependence of the ratio $\mathcal R(T_r)$
on the effective charge.
Here, the dashed and solid lines represent
the $1s$ and $2s$ sub-shells respectively.
And the different colors correspond to
different $Z_{\rm{eff}}$ values.
On one hand, the atomic effect caused
by the nuclear charge decreases as
$Z_{\rm{eff}}$ decreases. With
a smaller effective charge, one may expect the
electron wave function to move closer towards
the free electron case and the phase space
ratio to approach $1$. On the other hand,
the decrease in the nuclear charge
also reduces the phase difference between
the non-relativistic and relativistic final-state
wave functions. Such feature happens at
both the wave function level and the $K$-factor level.
Now, the origin of this suppression, which is
previously observed numerically in the literature 
\cite{Chen:2016eab,Bloch:2020uzh,Whittingham:2022wxc},
can become clear from the present analysis.

\subsection{Relativistic Effects in DM-Electron Scattering}
\label{sec:Relativistic Effects in Cross Section}

In the previous subsections, we have focused on
the general relativistic corrections to the
electron wave functions and the atomic $K$-factor.
In this subsection, we move to a more phenomenological
level and investigate how the relativistic atomic
effects affect the DM–electron scattering cross
section. For illustration, we take the scalar contact operator
$\mathcal{L}_{\text{eff}} \sim \bar \chi \chi \bar e e / \Lambda^2$
as an example.

\begin{figure}[t]
\centering
\includegraphics[width=0.68\textwidth]{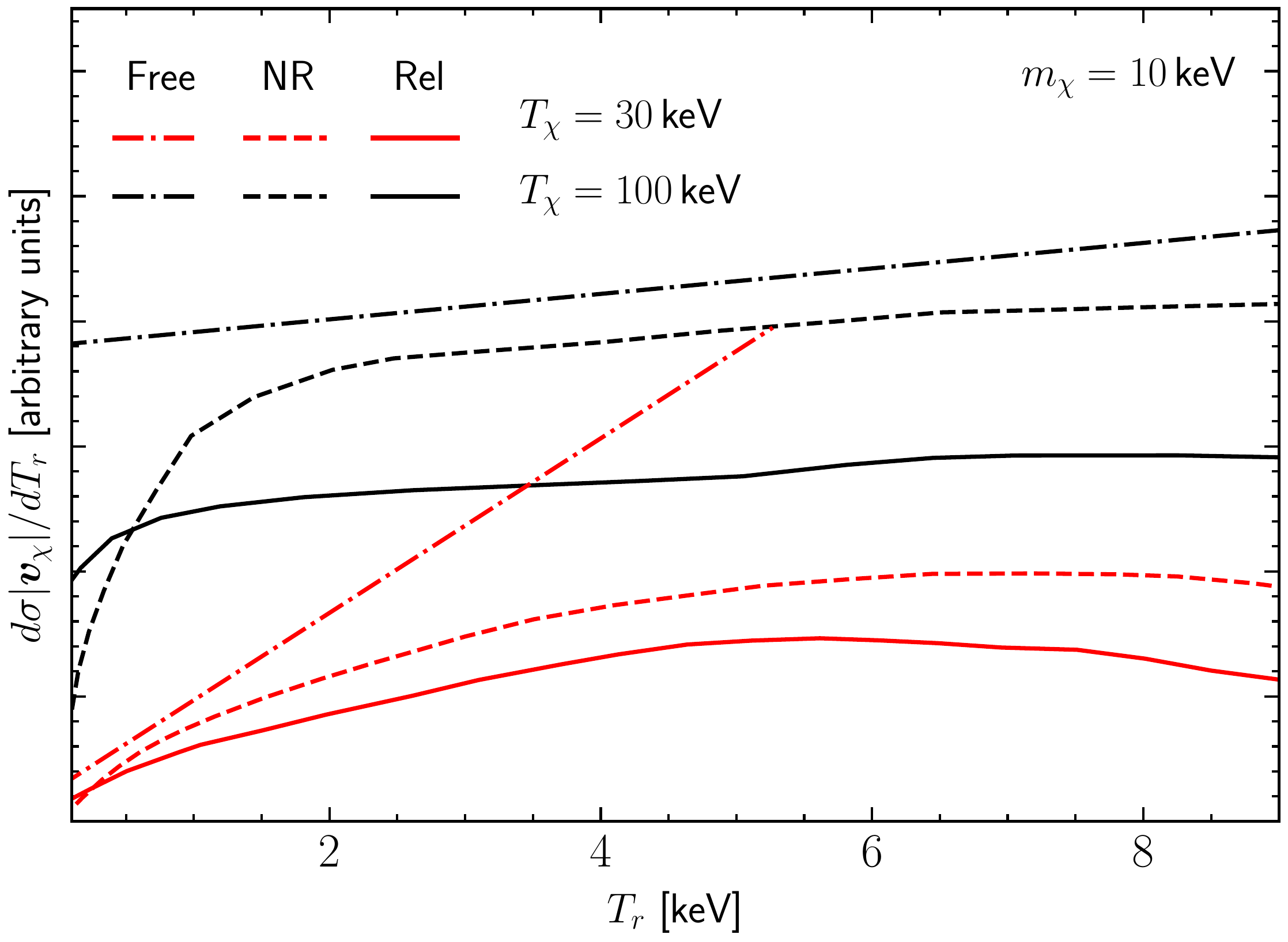}
\caption{
The differential cross section for DM scattering off a xenon atom when $m_\chi = 10\,$keV and incoming kinetic energies $T_\chi = 30$\,keV (red) and $100\,$keV (balck). The three different treatments with the free-electron approximation (dash-dotted), the non-relativistic (NR) atomic treatment (dashed), and the relativistic (Rel) formalism (solid) are shown for comparison.
}
\label{fig:dcs}
\end{figure}

To demonstrate that our method can truly avoid the inconsistency from
the kinematics as pointed out in \gsec{sec:Factorization},
we assign a light DM mass and large velocity. Such parameter
space can be naturally realized in the CRDM scenario
where the flux of relativistic light DM can be sizable
\cite{Ge:2020yuf,Xia:2021vbz,PandaX:2024pme}.
We show two benchmark cases
($m_\chi = 10\,\text{keV}$, $T_\chi = 30\,\text{keV}$) and
($m_\chi = 10\,\text{keV}$, $T_\chi = 100\,\text{keV}$)
with red and black curves in \gfig{fig:dcs}. 
For comparison and showing the impact of both the atomic
and relativistic effects, we show the differential scattering
cross sections in the free-electron approximation (dash-dotted),
the non-relativistic atomic treatment (dashed), and the
relativistic (solid) formalisms.

First, if we neglect the atomic effects altogether and make the approximation that DM scatters with a free electron at rest, the differential scattering cross section times velocity is given by,
\begin{equation}
    \frac{d \sigma_{\text{free}} |{\bm v_\chi}|}{dT_r}
=
    54 \times 
    \frac{1}{8 \pi \Lambda^4}
    \frac{(2 m_\chi^2 +  m_e T_r) (2 m_e + T_r)}{|{\bm p}_\chi| (m_\chi + T_\chi)}.
\label{cs_free}
\end{equation}
Here, the incoming DM with mass $m_\chi$ and velocity
$v_\chi$ carries an kinetic energy $T_\chi$ and
momentum $|{\bm p}_\chi|$ while $T_r$ is the electron
recoil energy. Denoting the number of electrons
in a Xenon atom, the prefactor $54$ is introduced
for direct comparison with those results below with
summation over all the atomic electrons to take the
atomic effects into consideration. 
For the free electron approach, the electron recoil
energy has an clear upper limit,
\begin{equation}
  T_r^{\text{max}}
=
  \frac {2m_e (T_\chi^2 + 2 m_\chi T_\chi)}
        {(m_\chi + m_e)^2+2 m_e T_\chi}.
\label{Trmax}
\end{equation}
For the case of $T_\chi = 30\,$keV (red)
in \gfig{fig:dcs}, the free-electron (red dash-dotted)
curve indeed stops at $T_r \simeq 5\,$keV.

Including atomic effects in the non-relativistic approximation,
the differential cross section times velocity is given by
\geqn{eq:dSigmadT_nr}, with the matrix element as \geqn{M_nr}. 
In the non-relativistic formalism, the trace of electron spinors
at rest gives $4 m_e^2$ while the DM counterpart is
$4 (2 m_\chi^2 - t/2)$ where $t = \Delta E^2 - |{\bm q}|^2$.
The nonrelativistic (NR) differential cross section
$d \sigma_{\rm NR} / d T_r$ then becomes,
\begin{equation}
    \frac{d\sigma_{\text{NR}} |{\bm v_\chi}|}{d T_r}
=
    \sum_{nl} \frac{2 (2l+1)}{16\pi^2 \Lambda^4 |{\bm p}_\chi| (m_\chi + T_\chi)} 
    \int_{{\bm q}_{\min}}^{{{\bm q}_{\max}}} 
    (4 m_\chi^2 + |{\bm q}|^2 - \Delta E^2)
    K_{nl}(|{\bm q}|, \Delta E) |{\bm q}| d|{\bm q}|.
\label{dcs_nr}
\end{equation}
Here, we have performed the summation over all electrons
in different quantum states. With approximations
$|{\bm q}|^2 \rightarrow 2 m_e T_r$,
$\Delta E^2  \rightarrow T_r^2$, and
$\int K_{nl}|{\bm q}| d|{\bm q}| \rightarrow 2\pi (m_e+T_r)$
in the free-electron limit, \geqn{dcs_nr} approximately reduces
to \geqn{cs_free} which can serve as a consistency check.
Similarly, once the relativistic atomic effects are
included, the scattering cross section
$d \sigma_{\rm R} / d T_r$ can be obtained
by inserting the DM spinor trace into \geqn{dcsWithK},
\begin{align}
  \frac{d \sigma_{\text{R}}|{\bm v_\chi}| } {d T_r} 
= \sum_{n \kappa}
  \frac{2|\kappa|} {16 \pi^2 \Lambda^4 |{\bm p}_\chi| (m_\chi + T_\chi)}
  \int_{{\bm q}_{\min}}^{{{\bm q}_{\max}}}
 (4 m_\chi^2 + |{\bm q}|^2 - \Delta E^2)
  K_{n \kappa} ({\bm q}, \Delta E) |{\bm q}| d|{\bm q}|.
\label{dc_r}
\end{align}
The relativistic differential cross sections are
shown as solid curves in \gfig{fig:dcs}.

Comparing with the free-electron approach (dash-dotted)
for $T_\chi = 30$\,keV,
the recoil spectrum extends to higher energy once the
atomic effects are included. This is because the initial
bound electron is not at rest and possesses a nontrivial
{\it fermi motion} in the atom. For a given momentum
transfer on the DM side, the target electron can be
scattered into a wider range of
final-state momenta. Quantitatively, the kinematics is
allowed as long as the expression under the square root
in \geqn{eq:q-limits} remains non-negative, namely
$T_r \lesssim T_\chi$.
For $T_\chi = 100\,$keV, the differential
cross section cut off calculated in the free-electron
frame is beyond the $T_r$ range shown in the figure
but the same feature still apply.

In addition, both the atomic effect and the relativistic
correction suppress the differential cross section.
This can be seen from the relative sizes of the
dash-dotted, dashed, and solid lines in both red and
black colors. For the $T_\chi = 100$\,keV benchmark,
the comparison among the three scenarios follows the
same behavior as phase space ratio discussed in the
previous subsection. The relativistic effects also
reduce the differential cross section by about
$30\% \sim 50\%$.
In the $T_r = 30\,$keV case, the suppression from atomic effects appears more pronounced. This is because, as $T_r$ increases and approaches $T_\chi$, the allowed integration range of $|{\bm q}|$ shown in \geqn{eq:q-limits} becomes narrower and the differential cross section decreases accordingly.

\section{Conclusion}
\label{sec:conclusion}

The atomic effects in the DM-bound electron scattering cannot be simply assumed as an additional atomic factor together with a free electron scattering matrix element, as this would render the scattering cross section theoretically inconsistent with negative event rate.
In this paper, we develop the formalism of bound-electron scattering within the framework of QFT where electron fields are directly quantized as bound and ionized states subject to the atomic Coulomb potential.
In the present work, the electron is never treated
as a free particle, but rather quantized in bound and ionized
states from the outset.
Based on these, we derive a different scattering cross section and the corresponding relativistic atomic form factor starting from the most fundamental definitions. 

To investigate the relativistic effects, we conducted comprehensive comparisons at various levels between the relativistic atomic form factor of scalar interactions and its non-relativistic counterpart. Our findings indicate that in the relativistic scenario, there are no significant differences in the structure of the atomic factor and the bound-state wave functions of electrons compared to the non-relativistic case. 
The primary disparity between the two lies in the final-state wave functions. As the electron recoil energy increases, the amplitude of the relativistic wave function decreases. Moreover, a larger effective charge leads to an increase in the phase difference between the relativistic and non-relativistic ionized wave functions. These relativistic effects result in a $30\% \sim 50\%$ reduction of the Xenon atomic factor shown in the phase space ratio. This reduction is correspondingly reflected in the DM–electron differential scattering cross section as well.
During the calculation of the atomic factor, we also discovered that by transforming the form of the confluent hypergeometric function, 
the time-consuming and challenging radial integration computations 
could be significantly sped up and improved in accuracy.

Although our calculation is illustrated with DM electron scattering,
it should also apply for the electron recoil signal with other neutral
incoming particles such as neutrino.

\section*{Acknowledgements}

The authors are supported by the National Natural Science
Foundation of China (12375101, 12425506, 12090060 and 12090064) and the SJTU Double First
Class start-up fund (WF220442604).
J.S. is supported by the Japan Society for the Promotion of Science (JSPS) as a part of the JSPS Postdoctoral Program (Standard) with grant number: P25018, and by the World
Premier International Research Center Initiative (WPI), MEXT, Japan (Kavli IPMU).
SFG is also an affiliate member of Kavli IPMU, University of Tokyo.
CYX is supported by the Fundamental Research Funds for the Central Universities (No. 24CX06048A).

\appendix
\section{The Simplification of $K$-Factor with Wigner-Eckart Theorem}
\label{Wigner_theorem}

To evaluate the $K$-factor in \geqn{eq:K0}, we first
simplify the position integration $\int d^3 {\bm r}$
and the triple summation $\sum_{\kappa' m'_j m_j}$ 
embedded in its definition.
The Fourier transformation kernel
$e^{i {\bm q \cdot \bm r}}$
can be expanded in terms of spherical harmonics,
$e^{i {\bm q \cdot \bm r}}
\equiv 
    \sum_L \sum_M 
    4 \pi i^L j_L (|{\bm q}| |\bm r|)
    Y^*_{LM} (\hat{\bm q}) Y_{LM} (\hat{\bm r})$ 
where $j_L$ is the Bessel function of the first kind 
while $\hat{\bm r}$ and $\hat{\bm q}$ denote the radius and
momentum directions, respectively. In this decomposition,
the position-dependent part can be defined as a tensor
operator $T_{LM}(\bm r, |\bm q|)
\equiv
    4 \pi i^L j_L (|{\bm q}| |\bm r|)
    Y_{LM} (\hat{\bm r})$ \cite{Sakurai:2011zz} . 
Additionally, by expressing the inner product of the
wave functions with operators in between as
$\braket{f|\mathcal O|i} \equiv \int d^3 {\bm r} \psi_f^\dagger \mathcal O \psi_i$,
the $K$-factor becomes,
\begin{equation}
  K_{n \kappa}^{S} ({\bm q}, \Delta E)
= 
    \frac{1}{2|\kappa|}
  \sum_{\kappa' m'_j m_j}
\left|
  \sum_{LM} Y^*_{LM} (\hat{\bm q})
  \braket{T_r \kappa' m'_j |
  \gamma^0_D T_{LM}(\bm r, |\bm q|)
  | n \kappa m_j}
\right|^2.
\label{Kbraket}
\end{equation}

The hydrogen-like atomic wave functions are
typically obtained by solving the Dirac equation
in the Dirac representation as shown later in
\gsec{sec:relSpinor}. As a result,
the subscript $D$ in the gamma matrix indicates it is also written in the Dirac representation to maintain the consistency.
Since $\gamma_D^0$
is block diagonalized, the Wigner-Eckart theorem \cite{Sakurai:2011zz}
can be applied to the tensor operator $T_{LM}$ to eliminate the dependence on the magnetic quantum number as, 
\begin{equation}
    \braket{T_r \kappa' m'_j
    | \gamma^0_D T_{LM}| n \kappa m_j}
=
    (-1)^{j'-m'_j}
    \begin{pmatrix}
      j' & L & j \\
      -m'_j& M & m_j
    \end{pmatrix}
    \left\langle T_r \kappa'
    \left\| \gamma^0_D
 T_{L} \right\|  n \kappa \right\rangle.
\end{equation}
The parenthesis above represents the Wigner-3j symbol and 
the inner product is a reduced matrix element \cite{Sakurai:2011zz}.
After performing the square in \geqn{Kbraket}, 
another Wigner-3j
symbol appears with a set of conjugate
indices $\bar L$ and $\bar M$. 
The summation over $m_j$ and $m'_j$ can be performed
using the orthogonality of Wigner-3j symbols \cite{Varshalovich:1988ifq},
\begin{equation}
    \sum_{m_j m'_j}
\left( 
    \begin{array}{ccc}
j & L & j' \\
m_j & M & m'_j
    \end{array}
\right)
\left(
    \begin{array}{ccc}
j & \bar L & j' \\
m_j & \bar M & m'_j
    \end{array}
\right)
=
    \frac{\delta_{L \bar L} \delta_{M \bar M}}
    {2 L + 1}.
\end{equation}
With these two $\delta$-functions, the summation over $\bar L$ and $\bar M$ can be performed.
Furthermore, the summation over $M$ is eliminated  
using $\sum_M |Y_{LM}(\hat q)|^2 = (2 L + 1)/4\pi$
\cite{Varshalovich:1988ifq}. Consequently, only the
summations over $\kappa'$ and $L$ remain in the
$K$-factor as, 
\begin{equation}
  K^S_{n,\kappa} ({\bm q}, \Delta E)
=
    \frac{1}{2|\kappa|}
  \sum_{\kappa' L}
    \frac{1}{4\pi} 
    \left|
    \left\langle T_r \kappa' 
    \left\| \gamma^0_DT_{L} \right\|   n \kappa
    \right\rangle
    \right|^2.
\label{fromfactor3}
\end{equation}
 
To calculate a reduced matrix element,
one needs to apply the 
Wigner-Eckart theorem once again to establish its connection with
an inner product of some typical states (wave functions).
Since the choice of states is
arbitrary, one can choose the simplest one with $m_j = 1/2$,
\begin{equation}
\left\langle T_r \kappa'
    \left\| \gamma^0_D T_{L} \right\|  
     n \kappa \right\rangle
=
    (-1)^{j' - \frac12}
    \begin{pmatrix}
      j' & L & j \\
      -\frac12 & 0 & \frac12 
    \end{pmatrix}^{-1}
\left\langle
  T_r \kappa' \left. \frac12 
  \right| \gamma^0_D T_{L 0}
  \left| n \kappa \frac 1 2 \right.
\right\rangle.
\label{WE1/2}
\end{equation}
After making the above substitutions, 
the final expression of the atomic factor
for scalar interaction becomes,
\begin{equation}
  K^S_{n,\kappa} (\Delta E, {\bm q})
=
    \frac{1}{2|\kappa|}
  \sum_{\kappa'} \sum_{L}
    \frac{1}{4 \pi} 
    \begin{pmatrix}
      j' & L & j \\
      -\frac12 & 0 & \frac12 
    \end{pmatrix}^{-2}
    \left|
\left\langle
  T_r \kappa' \left. \frac12 
  \right| \gamma^0_D T_{L 0}
  \left| n \kappa \frac 1 2 \right.
\right\rangle
    \right|^2.
\label{kernel3prime}
\end{equation}
After simplification, the original triple summation reduces to a double summation, and the inner product of initial and final wave functions now incorporates only states with magnetic quantum number $m_j = 1/2$.

\section{A Simplified Numerical Procedure for Computing the $K$-Factor}
\label{numerical}

In the electron wave functions, 
this oscillatory behavior stems mainly from the confluent hypergeometric functions $_1F_1$ in \geqn{Rel_Ionized_L}. Usually, the $_1 F_1$ function is expanded as a power series.
It contains finite terms
for the bound states,
\begin{equation}
    {}_1F_1 (-n' , 2\gamma +1, \frac{2 Z_{n \kappa} |\bm r|}{N})
    =
    \sum_{s=0}^{n'}
    \frac{(-n')^{(s)}}{(2\gamma + 1)^{(s)} s !} 
    \left( \frac{2 Z_{n \kappa} |\bm r|}{N} \right)^s.
    \label{F_expansion}
\end{equation}
The calculation rule defined by the 
superscript $(s)$ is $x^{(0)}=1$ and 
$x^{(s)}=x(x+1)(x+2) \cdots(x+s-1)$.
However, the series expansion of ${}_1 F_1$
in the ionized final-state wave function contains an
infinite number of terms \cite{Andrews_Askey_Roy_1999},
making the integration over radius $r$ within an infinite
range $(0, + \infty)$ very difficult to be calculated. 

To address this, we write the confluent hypergeometric
function of the final-state wave function in the
integral form \cite{Andrews_Askey_Roy_1999},
\begin{equation}
    {}_1F_1(a, b, z)
    \equiv
    \frac{\Gamma(b)}{\Gamma(a) \Gamma(b-a)} 
    \int_0^1 \mathrm{e}^{z u} u^{a-1}(1-u)^{b-a-1} d u,
\label{F2integral}
\end{equation}
which holds for $
    \mathrm{Re} (b) > \mathrm{Re} (a) > 0$ with $a \equiv \gamma +1 - i \nu \, (a \equiv \gamma - i \nu)$ for the first (second) $_1 F_1$ function in \geqn{Rel_Ionized_L} and 
$b \equiv 2 \gamma + 1$.
In a Xenon atom, $\gamma$ is always positive such
that the above requirement can be satisfied.
The radial integration $\mathcal R$ now contains the integration over both radius $|\bm r|$ and the new parameter $u$ as, 
\begin{equation}
    \mathcal R_{PP,QQ}
    \supset    
    \int_0^1 d u \,
    u^{a^*-1}(1-u)^{b^* - a^* -1} 
    \int d |\bm r| \,
    |\bm r|^{\alpha} 
    \mathrm{e}^{- \beta |\bm r|} 
    j_L(|\bm q| |\bm r|) .
\end{equation}
Since the final-state wave function appears in the
inner product in the form of a complex conjugate,
the exponential power $a^*$ ($b^*$) also takes a
complex conjugate. While the final-state wave function
$P_{T_r \kappa}$ ($Q_{T_r \kappa}$) in
\geqn{Rel_Ionized_L} contributes the part as function
of $u$, the radial exponential $e^{z u}$ in
\geqn{F2integral} combines with the $e^{- Z_{n \kappa} |\bm r| / N}$
factor of $P_{n \kappa}$ ($Q_{n \kappa}$)
in \geqn{Rel_Bound_Wave_L}. Together with
the $e^{i |\bm p| |\bm r|}$ in \geqn{Rel_Ionized_L}, the
exponential term receives
$\beta \equiv Z_{n \kappa} /N + i |\bm{p}| - 2 i |\bm{p}| u$.
For the power series of $|\bm r|$, $P_{T_r \kappa}$
($Q_{T_r \kappa}$) contains a factor
$(- 2 i |\bm p| |\bm r|)^{\gamma_I}$ and
$P_{n \kappa}$ ($Q_{n \kappa}$) has
$(2 Z_{n \kappa} |\bm r| / N)^{\gamma_B}$. With the $_1F_1$ functions
in \geqn{Rel_Bound_Wave_L} expanded to give $|\bm r|$
series, the $|\bm r|$ power series finally receive
$\alpha \equiv \gamma_B + \gamma_I + s$. The powers
$\gamma_B$ and $\gamma_I,$ come from the bound and
ionized state wave functions in \geqn{Rel_Bound_Wave_L}
and \geqn{Rel_Ionized_L}, respectively, while $s$
comes from the series expansion of the ${}_1 F_1$
function in the initial-state wave function \geqn{F_expansion}.

Regardless of the forms of $\alpha$ and $\beta$,
the integral over $|\bm r|$ in such a form can be evaluated
analytically by the Gauss Hypergeometric Function
${}_2 F_1$ \cite{Andrews_Askey_Roy_1999},
\begin{equation*}
    \int d |\bm r|
    |\bm r|^\alpha \mathrm{e}^{-\beta |\bm r|} 
    j_L(|\bm q| |\bm r|) 
    = 
    \frac{\sqrt{\pi} q^L}{2^{L+1} \beta^{\alpha+L+1}}
    \frac{\Gamma(L+\alpha+1)}{\Gamma\left(L+\frac{3}{2}\right)}
    {}_2 F_1\left(\frac{L+\alpha+1}{2}, \frac{L+\alpha+2}{2}, L+\frac{3}{2},-\frac{q^2}{\beta^2}\right) ,
\end{equation*}
when $\mathrm{Re}(L+\alpha)>-1$. 
As a result, the original integration over radius $\mathcal R_{PP,QQ}$ in the range of $|\bm r| \subset (0, + \infty)$ is now reduced to a sum of integrals over $u$ in the range of $u \subset [0,1]$.
This reduction significantly enhances the efficiency of the numerical calculation of atomic factors. 
The numerical result of the $K$-factor depends on both the range of the sum over angular-momentum quantum numbers and the numerical accuracy of the integrations.

\providecommand{\href}[2]{#2}\begingroup\raggedright\endgroup

\end{document}